\documentclass[aps,prl,twocolumn,showpacs,superscriptaddress]{revtex4-1}

\usepackage[utf8]{inputenc}
\usepackage{verbatim}

\usepackage{graphicx,amsmath,amssymb,xspace}

\usepackage{xcolor}
\usepackage{physics,mathtools}

\begin{document}

\title{Exact spectral function of a Tonks-Girardeau gas in a lattice}

\author{J. Settino}
\affiliation{Dipartimento di Fisica, Universit\`a della Calabria, I-87036
Arcavacata di Rende (CS), Italy}
\affiliation{CNR-SPIN, I-84084 Fisciano (Salerno), Italy}
\author{N. Lo Gullo}
\affiliation{QTF Centre of Excellence, Turku Centre for Quantum Physics,
Department of Physics and Astronomy, University of Turku, 20014 Turku, Finland}
\author{F. Plastina}
\affiliation{Dipartimento di Fisica, Universit\`a della Calabria, I-87036
Arcavacata di Rende (CS), Italy}
\affiliation{INFN, gruppo collegato di Cosenza}
\author{A. Minguzzi}
\affiliation{Univ. Grenoble-Alpes, CNRS, LPMMC, 38000 Grenoble, France}

\begin{abstract}
  The single-particle spectral function of a strongly correlated system is an essential ingredient to describe its dynamics and transport properties. 
  We develop a general  method to calculate the exact spectral function of a strongly interacting one-dimensional Bose gas in the Tonks-Girardeau regime, valid for any type of confining potential,  and apply it to bosons on a lattice to  obtain  the full spectral function, at all  energy and momentum scales.  We find that it  displays three main singularity lines. The first two can be identified as the analogs of Lieb-I and Lieb-II modes of a uniform fluid; the third one, instead, is specifically due to the presence of the lattice.  We show that the spectral function displays a power-law behaviour close to the Lieb-I and Lieb-II singularities, as predicted by the non-linear Luttinger liquid description, and  obtain the exact exponents. In particular,  the Lieb-II mode shows a divergence in the spectral function, differently from what happens in the dynamical structure factor, thus providing a route to probe it in experiments with ultracold atoms.
\end{abstract}
\maketitle

\paragraph*{Introduction.}
The dynamics of interacting many-body systems is a very active research field. Ultracold atomic gases offer an ideal experimental platform for such studies, thanks to the possibility of choosing particle statistics and of tuning interactions, geometry and dimensionality of the system. Astonishing experimental advances in realizing, controlling and measuring such systems to high precision allow to address  fundamental open questions, such as  the description of the arbitrarily long-time dynamics and the behavior of one-dimensional systems with broken integrability.
In this context the Tonks-Giradeau (TG) gas \cite{Girardeau1960,Bijl1937} deserves special mention. It is a system of strongly correlated one-dimensional bosons, with infinite repulsive interaction. This regime has been experimentally achieved with ultracold atoms \cite{Olshanii1998,Moritz2003,Kinoshita2004,Paredes2004} allowing to  study correlation and many-body effects \cite{Kinoshita2006,VanAmerongen2008,Palzer2009,Jacqmin2011,Meinert2017,Wilson2020}.
Thanks to the possibility of describing the TG  many-body wavefunction by an exact solution,  several facets have been deeply investigated:  one-body density matrix  \cite{Lenard1966,Vaidya1979,Jimbo1980,Girardeau2001,Papenbrock2003,Forrester2002,Forrester2003,Castin2004,Rigol2005,Yukalov2005,Vignolo2013,Garcia-March2015,Xu2017,Settino2017,Atas2017,Colcelli2018,Lang2017,Brun2017},  momentum distribution \cite{Lenard1964,Girardeau2001,Minguzzi2002,Lapeyre2002,Rigol2004,Minguzzi2005,Rigol2006,Pezer2007,Deng2008,Vignolo2013,Xu2017,Settino2017}, and non-equilibrium  properties~\cite{Das2002,Berman2004,Minguzzi2005,Rigol2006,Rigol2007,Kormos2014,Boumaza2017,Bastianello2017,YagoMalo2018,Mikkelsen2018}.

\begin{figure}[!ht]
 \centering
\includegraphics[width=1\linewidth]{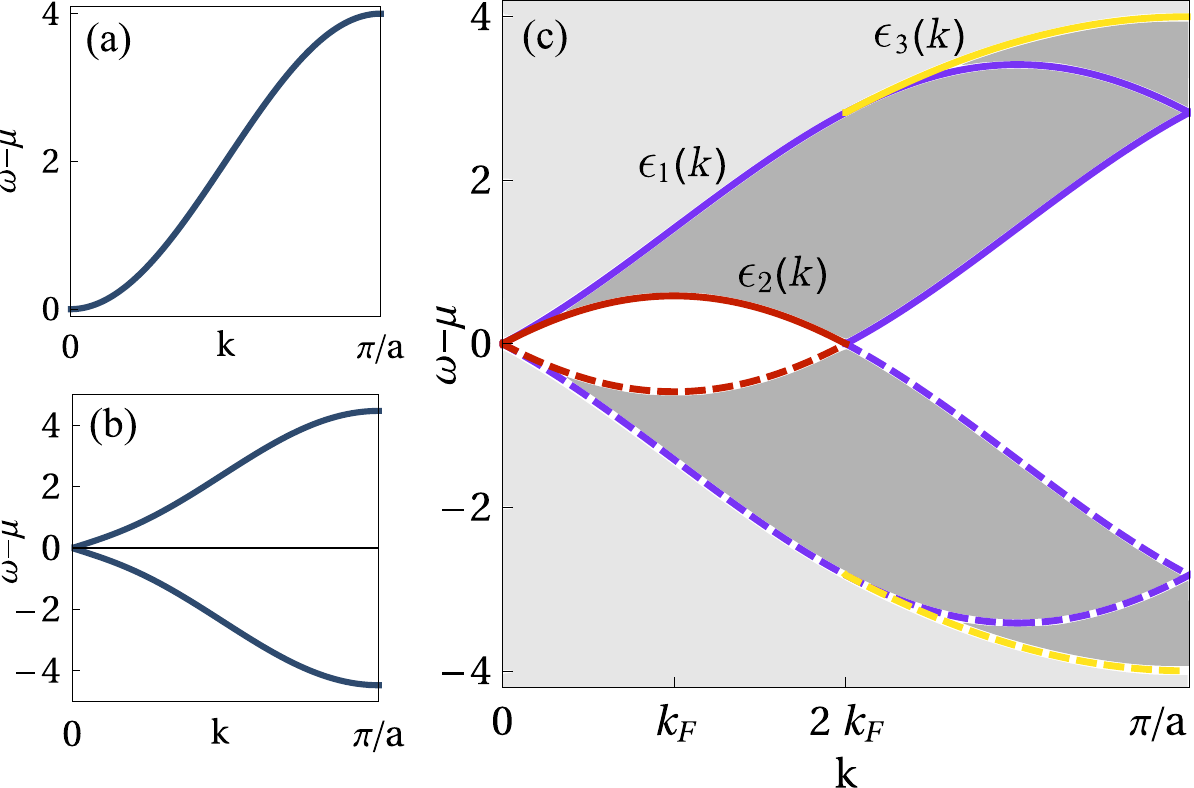}
\begin{minipage}[!ht]{0.31\linewidth}
        \includegraphics[width=\linewidth]{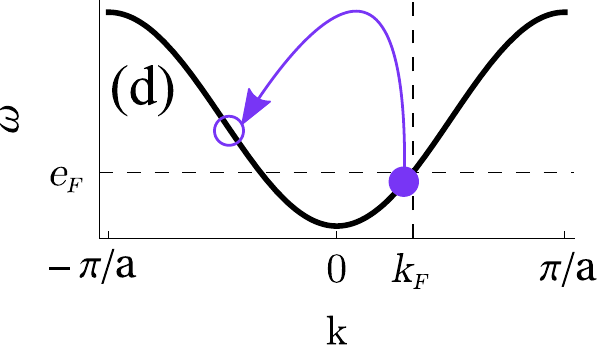}
    \end{minipage}
    ~ 
    \begin{minipage}[!ht]{0.31\linewidth}
        \includegraphics[width=\textwidth]{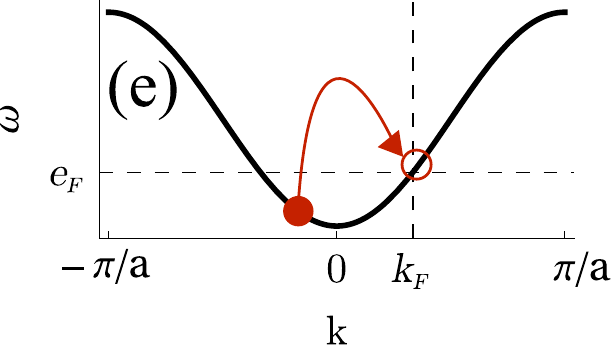}
    \end{minipage}
    ~ 
    \begin{minipage}[!ht]{0.31\linewidth}
        \includegraphics[width=\textwidth]{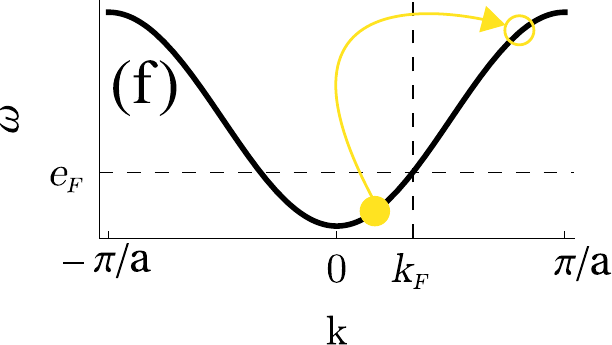}
    \end{minipage}
    \caption{(Color online) \label{fig:Sketch}Sketch of the expected spectral function of a  Bose fluid on a lattice in the $(k,\omega)$  plane for (a) zero, (b) weak,  and (c) strong interactions. In this work we provide an exact solution  for the regime of  infinitely strong interaction, which describes generically the case (c). The shaded areas indicate the regions where the spectral function is non-zero. Panels (d), (e) and (f) show the processes that give rise to the singularity lines $\epsilon_{1,2,3}(k)$.}
    \label{fig:scheme}
\end{figure}

A primary quantity in many-body physics is the spectral function. It embodies information about the accessible energy states and their distribution in momentum space. Its knowledge is of pivotal importance in the characterization of the dynamical properties of the system. Specifically, it allows to compute the signal of either angle-resolved photoemission spectroscopy (ARPES), or momentum-resolved stimulated Raman spectroscopy, which have recently been performed with cold atoms platforms ~\cite{Damascelli2004,Stewart2008,Dao2009,Volchkov2018,Bohrdt2018}. Moreover, it gives access to the transmission coefficient of the system through which transport properties can be assessed by using the Landauer-B\"uttiker formula~\cite{Stefanucci2010,Tuovinen2014,Tuovinen2017,Talarico2020}.
As represented in Fig.\ref{fig:scheme} (a), for non-interacting and weakly interacting bosons, the spectral function consists of a sharp distribution along the energy dispersion relation of particles or quasiparticles (Bogoliubov excitations). Beyond-mean-field effects yield a broadening of the spectral function, due to a continuum of possible excitation processes allowed by particle correlations. For one-dimensional fluids in absence of confinement, the non-linear Luttinger liquid theory predicts the shape of the spectral function near the excitation singularities, based on the knowledge of their position  \cite{Khodas2007,Imambekov2008,Pereira2008,Imambekov2009,Imambekov2009a,Kamenev2009,Imambekov2012,Ristivojevic2014,Markhof2016,Campbell2017}. Other than those, only few studies have been devoted to the understanding of the spectral function.  Indeed, exact Bethe Ansatz calculations  are challenging due to  evaluation of  form factors and are restricted to the integrable case of uniform systems \cite{Kozlowski2011}, while numerical calculations of correlation functions are computationally demanding since they require to follow the many-body dynamical evolution at long times.

In this work, we provide an alternative route to the calculation of the spectral function of strongly correlated one-dimensional bosons using the exact Girardeau many-body wavefunction. Specifically, we present an efficient method  to compute it in terms of a functional of single-particle states and apply it to study the spectral function of lattice bosons. For this case, we identify three main singularity lines, characterize their power-law decaying spectral weight, and compare our results to earlier ones obtained by means of the non-linear Luttinger liquid theory. Besides this specific example, we emphasize that our method is completely general, and valid for any confining potential. It provides access to the spectral function at all energy and momentum scales, thus allowing direct comparison with current state-of-the-art experiments. 
Our results open up the possibility of studying the long-time dynamics of a TG in the non-equilibrium Green's functions framework~\cite{Stefanucci2010,Talarico2019} and specifically to investigate the competing role of strong correlations, external trapping potential and baths \cite{LoGullo2015,LoGullo2016,Settino2020,Talarico2020}.

\paragraph*{Model and physical quantities.}
We consider a gas of $N$ interacting bosons  at zero  temperature, tightly confined in a one-dimensional atomic waveguide.  Its Hamiltonian reads:
\begin{equation}
 \label{eq:ham}
 \hat H =\sum_{i=1}^N\left[
      -\frac{\hat p_i^2}{2m}
      +V(\hat x_{i})\right]
      +g \sum_{i<j=1}^N \delta (\hat x_{i}- \hat x_{j})\;
\end{equation}
with $ V(x)$ being an external potential and $g$ the coupling strength for the contact interaction in one-dimension \cite{Olshanii1998}. The integrable Lieb-Liniger model \cite{Lieb1963,Lieb1963a,Yang1969,Jimbo1980,Kozlowski2011,Cazalilla2011} is recovered in the uniform case $V(x)=0$.

The TG regime corresponds to  $g\rightarrow\infty$. As pointed out in \cite{Girardeau1960},  in this limit bosons are  impenetrable  and the effect of interactions can be embedded in the cusp condition on the many-body wavefunction
\begin{equation}
  \label{eq:ccTG}
  \Psi_B(x_1,x_2,\dots,x_N)=0\quad \mbox{if} \quad x_i=x_j\;,
\end{equation}
for $i\neq j$ and $1\leq i\leq\ j\leq N$.
An exact solution for $ \Psi_B$ is obtained  \cite{Girardeau1960} by mapping  the strongly interacting boson gas into a system of non-interacting  fermions in the same external potential with wavefunction  $\Psi_F$,
\begin{equation}\label{eq:mapping}
\Psi_B= {\cal A} \Psi_F\mbox{, ~~ with ~~}  {\cal A}=\!\!\!\!\!\prod_{1\leq i < j\leq N} \mbox{sign}(x_j-x_i)\;.\\
\end{equation}
One consequence of this mapping is that all local quantities,
e.g.~the spatial density of particles and the dynamic structure factor, coincide with those of a non-interacting fermionic gas. The difference between these two systems emerges when looking at non-local quantities, such as the  momentum distribution, which displays typical boson features, as a main peak at zero momentum \cite{Forrester2003}, as well as effects of short range interactions in the high-momentum tails \cite{Vignolo2013,Olshanii1998}. One important example of a non-local quantity is the single-particle spectral function,
\begin{equation}\label{eq:SF}
 A(k,\omega) = - \frac{1}{\pi} \text{Im} G^R(k,\omega)
\end{equation}
obtained as the Fourier transform of the retarded Green's function
$G^R(x,t;y,t')=\theta(t-t')  \left[ G^>(x,t;y,t')-G^<(x,t;y,t') \right]$. Here, $G^<(x,t;y,t')= -\imath \expval{  \hat \Psi^\dagger(y,t') \hat \Psi (x,t) }$,  and  $G^>(x,t;y,t')= -\imath\expval{ \hat \Psi (x,t) \hat \Psi^\dagger(y,t') }$
are  the lesser and greater Green's functions, typically employed in non-equilibrium theory \cite{Stefanucci2010},  with $\Psi (x,t)$ and $\Psi^\dagger(x,t)$ being the bosonic field operators.
The spectral function gives to the transition amplitude for exciting a particle (hole) with energy $\omega$ ($-\omega$) and momentum $k$.
It is worth noting that, in the bosonic case, it can be negative~\cite{Stefanucci2010}, therefore losing its probability-density interpretation.
As it is customary, we  analyze  separately the Fourier transforms of the lesser and greater Green's functions, corresponding to the probability-density for a particle (hole) to be excited (filled) at a given energy-momentum pair.

\paragraph*{Single-particle Green's functions for the TG gas.}
Employing Eq.(\ref{eq:mapping}) together with the definition of $G^{<,>}$, we  obtain  an explicit expression for the Green's functions in terms of one-dimensional integrals of single-particle orbitals (see Appendices ~\ref{app:lessgreat} and ~\ref{app:finallg} for the derivation). These expressions, which constitute our main result, are the following
\begin{subequations}
\begin{equation}\label{eq:lesserTG}
\imath G^< (x,t,y,t')=  \text{Det}[\textbf P (x,t)\textbf P (y,t')|_{{\boldsymbol \eta}{\boldsymbol \eta}}] a^<(x,t,y,t')
\end{equation}
\begin{equation}\label{eq:greaterTG}
 \imath G^> (x,t,y,t')= \text{Det}[\textbf P (y,t')\textbf P (x,t)|_{{\boldsymbol \eta}{\boldsymbol \eta}}] a^>(x,t,y,t')
\end{equation}
\end{subequations}
with
\begin{subequations}
\begin{align}
 a^<(x,t,y,t') &= {\boldsymbol \phi(x,t)_{{\boldsymbol \eta}}^T} \  {[{\textbf P}(x,t) {\textbf P}(y,t')]^{-1 T}}|_{{\boldsymbol \eta}{\boldsymbol \eta}} \   {\boldsymbol \phi^*(y,t')}_{{\boldsymbol \eta}}\\
  \begin{split}
  a^>(x,t,y,t') &= \boldsymbol \phi(y,t')^\dagger \boldsymbol \phi(x,t) -[\boldsymbol \phi(y,t')^\dagger \textbf P (x,t)]_{\boldsymbol \eta}\ \\& [\textbf P (y,t')\textbf P (x,t)]^{-1}|_{{\boldsymbol \eta}{\boldsymbol \eta}} \ [\textbf P (y,t') \boldsymbol\phi(x,t)]_{{\boldsymbol \eta}}
  \end{split}
  \end{align}
\end{subequations}
Here,  $\boldsymbol \phi(x,t)= [\phi_1(x,t),\dots,\phi_M(x,t)]^T$ is the column-vector of the single-particle orbitals, with $M$ being the single-particle Hilbert space dimension (or truncation dimension). The central quantity entering the equations above is the matrix ${\textbf P}$, with matrix elements $\text {P}_{lm}(x,t)= \int_{-\infty}^{\infty} \text{sign}(x-\bar x)\phi_l(\bar x)\phi^*_m(\bar x) d \bar x =\delta_{l,m} - 2 \ e^{-\imath t (\epsilon_l - \epsilon_m)} \int_{x}^{\infty} \phi_l(\bar x)\phi^*_m(\bar x) d \bar x$, where $\epsilon_l$ is the energy level corresponding to the orbital  $\phi_l(x)$ and $\boldsymbol \eta$ is the vector of integers that identify the single particle states that form the many-body eigenstate of the TG Hamiltonian. Henceforth, we work with the ground state and therefore $\boldsymbol \eta=1,\dots,N$. 
For fixed space and time coordinates, products between matrices $\textbf P$ or $\boldsymbol \phi$ run over the whole single-particle Hilbert space and are then projected through the indices ${\boldsymbol \eta}$.

From the above expressions, one readily  recovers the known results in the non-interacting fermion limit. It suffices to replace  $\text{sign}(x-y)$ with $1$, which gives $P_{l,m}(x,t)=\delta_{l,m}$, and hence
$G^<_{F} (x,t,y,t') = \imath \sum_{\boldsymbol \eta} e^{ \imath \epsilon_i t'}\phi^*_i (y) \phi_i (x) e^{- \imath \epsilon_i t} $ and
$G^>_{F} (x,t,y,t') = -\imath \sum_{\bar{\boldsymbol \eta}}  e^{ \imath \epsilon_i t'}\phi^*_i (y) \phi_i (x) e^{- \imath \epsilon_i t}$, which
are the  single particle Green's functions for a gas of $N$ non interacting fermions in the state $\boldsymbol \eta$ \cite{Stefanucci2010}.
Most importantly, our expression for the lesser Green's function in Eq.~(\ref{eq:lesserTG}) contains as a limiting case the result derived by Pezer and Buljan \cite{Pezer2007} for the  one-body  density matrix at equal times, $\rho(x,y)=-i G^<(x,t;y,t)$, for which it provides a generalization for $t\ne t'$.  Ref.~\cite{Pezer2007} is  one of the most efficient algorithms to evaluate the one-body density matrix, and allowed to perform several studies on the properties of the TG gas. Quite remarkably, the formal analogy between Eq.(\ref{eq:lesserTG}) and the one in ~\cite{Pezer2007}  implies that the calculation of the two-time Green's function requires a similar computational effort as the equal-time one.

As an application of the method,  we employ  Eqs. (\ref{eq:lesserTG}) and (\ref{eq:greaterTG}) to obtain the spectral function of a Tonks-Girardeau gas on a  lattice. In this case, we calculate the single particle orbitals $\phi_\ell(j)$ and energy levels $\epsilon_\ell$  of the Hamiltonian  $\hat{H}=-J\sum_{j=1}^{N_s-1} \hat{b}_j^\dagger \hat{b}_{j+1}+{\text{h.c.}}$, with $b_j$, $b^\dagger_j$ being the lattice boson operators and $N_s$  the number of lattice sites.  We impose open boundary conditions to model an additional box trap confinement.

\begin{figure}[!ht]
\centering
\includegraphics[width=1\linewidth]{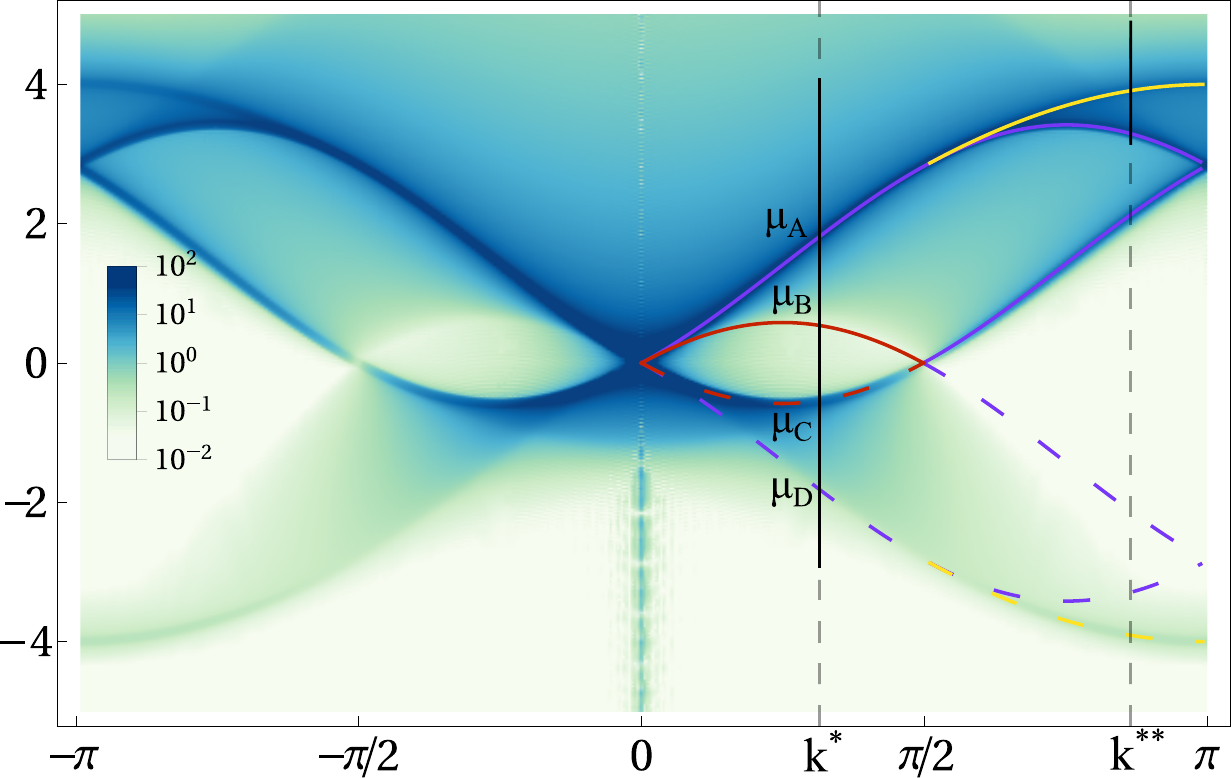}
\caption{(Color online) Spectral function of the TG gas on a lattice in the $(k, \omega)$ plane.
      Violet, red and yellow lines mark the excitation singularity lines  $\epsilon_1(k)$, $\epsilon_{2}(k)$ and $\epsilon_{3}(k)$, respectively, which  correspond to the elementary processes depicted in the bottom panels of Fig. \ref{fig:Sketch}, from left to right, respectively. Parameters used in the calculation: $N_s=256$, $N=64$.}
  \label{fig:SFLat}
  \end{figure}

\paragraph*{Spectral function of the TG gas.} 
Our results for the spectral function of the TG gas are presented in  Fig.~\ref{fig:SFLat}. The $\omega \ge (\le) \epsilon_F$ part of $A(k,\omega)$ comes from the greater (lesser) Green's functions,  Eqs. \eqref{eq:lesserTG} and \eqref{eq:greaterTG},  $\epsilon_F$ being  the Fermi energy of the mapped Fermi gas. The spectral function is characterized by three main singularity lines, denoted as  $\epsilon_1(k)$, $\epsilon_{2}(k)$ and $\epsilon_{3}(k)$, corresponding to specific excitation processes (see the bottom panels of Fig.\ref{fig:Sketch}). The first two  are analogous to those predicted by Lieb and Liniger~\cite{Lieb1963a} for a homogeneous Bose gas.
In detail, $\epsilon_1(k)$ corresponds to a Lieb-I (particle-like) excitation process, where a particle from the highest occupied state, with momentum $k_F$, is promoted to a generic non-occupied state with momentum $k_F+k$ (Fig.~\ref{fig:Sketch}d);  $\epsilon_{2}(k)$ corresponds to a Lieb-II (hole-like) excitation, from an occupied state with momentum $k_F-k +2\pi/L $, to the first unoccupied state with momentum $k_F+2\pi/L$ (Fig.~\ref{fig:Sketch}e). As in the homogeneous system, the spectral function vanishes in the regions $|\omega-\epsilon_F|<\epsilon_{2}(k)$, where no excitation is kinematically allowed due to the underlying Fermi-sphere structure of the ground state. This condition holds also for arbitrary interaction  \cite{Imambekov2012}. In addition to these two dispersion lines, we identify a third one, $\epsilon_{3}(k)$,  generated by the symmetric excitation of a particle from an occupied state at momentum $k/2$ to a free one with momentum $\pi/a - k/2$ (Fig.~\ref{fig:Sketch}f). This process can occur because of  the lattice induced periodicity of single-particle dispersion, $\xi(k)=-2J\cos(k a)$, which changes curvature  at $k\sim \pm \pi/a$,  and has no analogue in the homogeneous case.

\begin{figure*}[t b]
\includegraphics[width=\textwidth]{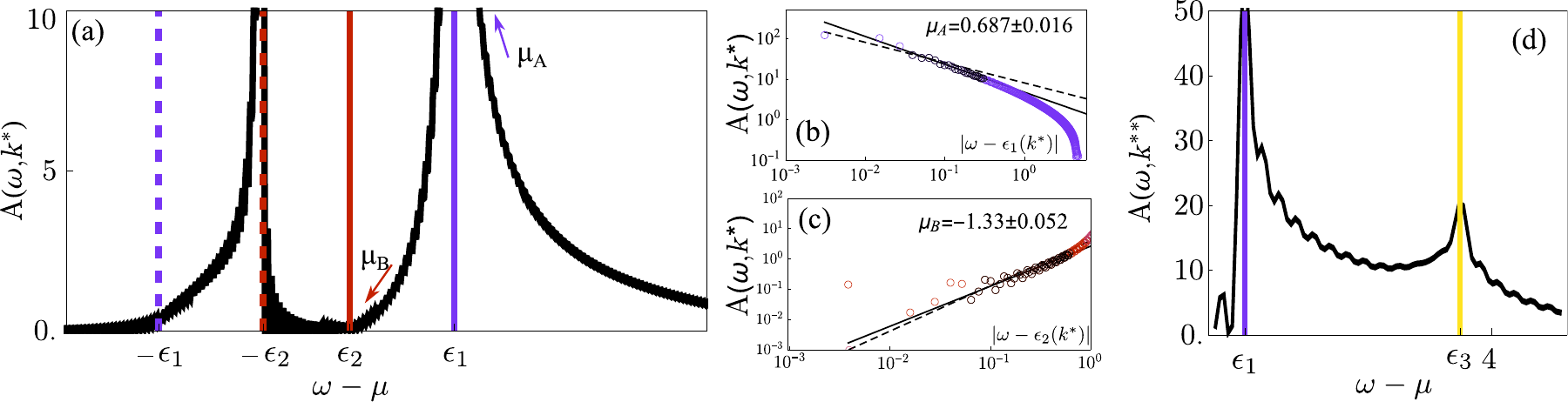}
\caption{\label{fig:SFCut}
  (a): Cut of the spectral function for the TG gas in a lattice (Fig. \ref{fig:SFLat}) for $k^*=1.0$. It shows power law behaviour in the vicinity of each depicted line ($\pm \epsilon_1(k^*), ~ \pm \epsilon_2(k^*)$).
   (b) and (c): Power law behaviour of the SF, fitted with $ c (\omega - \epsilon_{1,2}(k^*))^{-\mu}$, together with the fitted values of the exponents. d): Another cut of the spectral function at $k^*=2.71$, close to $\epsilon_3(k^{**})$. Vertical lines correspond to the peaks at  $\omega-\epsilon_F=\epsilon_1(k^{**})$ and  $\omega-\epsilon_F=\epsilon_3(k^{**})$. }
  \end{figure*}

We next analyze the behaviour of the spectral function near each excitation branch (see  Fig.~\ref{fig:SFCut}). For the homogeneous Bose gas, the non-linear Luttinger liquid theory predicts  a  power-law behavior, $A(k,\omega)\propto |\omega-\epsilon_j(k)|^{-\mu_j}$, for the spectral function  near each excitation line $\epsilon_j(k)$, with power law exponent $\mu_j$. These exponents, one for each excitation branch, can be calculated via the mobile impurity theory \cite{Imambekov2008,Imambekov2009,Imambekov2009a,Imambekov2012,Campbell2017}.
For the homogeneous TG gas, their predicted values are $\mu_A=1/2$, $\mu_B=-3/2$, $\mu_C=1/2$ and $\mu_D=-3/2$, giving rise to diverging non-analyticities for positive exponents, and converging
non-analyticities for negative ones. %
With our exact calculation, we find power-law behaviours also in the presence of the lattice, close to the singularities  $\omega - \epsilon_F =- \epsilon_2(k)$ and $\omega - \epsilon_F = \epsilon_1(k)$, while we obtain vanishing non-analyticities at $\omega - \epsilon_F = \epsilon_2(k)$ and  at $- \epsilon_{1}(k)$.
This behavior is illustrated in Fig.~\ref{fig:SFCut}, where we display various cuts of the spectral function at fixed $k$. There, we also provide the values of the power-law exponents obtained by fitting the exact  TG spectral function (see Appendix ~\ref{app:pwrlaw} for details). The exponents are close, but not exactly coinciding  with the predicted values $\mu_A,...\mu_D$  for a Lieb-Liniger gas \cite{Imambekov2008}. This is expected because we consider a lattice system, though at relatively low filling. Furthermore,  our exact method goes beyond the approximations employed in \cite{Imambekov2008}: a similar renormalization effect of a power-law exponent is found for the so called Fermi edge singularity, when comparing an exact numerical calculation with a perturbative one \cite{Mahan1967,Nozieres1969a,Sindona2013,Sindona2015}. We also notice that the approximate power-law description holds in a very narrow interval close to the singularity \cite{Imambekov2009a}.

As mentioned above, he spectral function also shows a marked structure at  $\omega - \epsilon_F =\pm \epsilon_{3}(k)$, due to the presence of the lattice. However, this is not expected to give rise to a divergence, see Ref. \cite{Pereira2008}.
We stress that our calculation is exact  (within the numerical accuracy) at all energy and momentum scales. For instance, we have checked that the momentum distribution, obtained by integration over all frequencies of Im$G^<(k,\omega)$  displays the expected high-momentum $k^{-4}$ tails \cite{Minguzzi2002,Olshanii2004} not captured by the Luttinger-liquid description.
Finally, we notice that, due to correlation effects, the spectral function is not vanishing for $|\omega-\epsilon_F|>4J$ (see Fig.~\ref{fig:SFCut} (d)), as it is the case for non-interacting bosons, where the maximum allowed energy exchange corresponds to moving a particle from the bottom to the top of the single-particle energy spectrum (Fig.~\ref{fig:Sketch} (b). Here, instead, an infinity of high-energy levels are involved in the Green's functions (see again  Eqs.~\eqref{eq:lesserTG} and \eqref{eq:greaterTG}).

\paragraph*{Conclusions.}
We have obtained an exact analytical expression for  the lesser and greater Green's functions of a Tonks-Girardeau gas in terms of one-body integrals of single-particle orbitals. Our method applies to any form of external potential and  allows for efficient numerical calculations. We have used these expressions to evaluate the spectral function of the Tonks-Girardeau gas in a lattice. For this case, we have identified three singularity lines, two of which are typical of homogeneous Bose gases, while the additional one is due to the lattice confinement. Close to the diverging singularities, the spectral function shows power-law behaviors, as predicted by the non-linear Luttinger liquid theory. Our  description allows to obtain the exact power-law exponents, as well as the exact behaviour at all energy and momentum scales. The spectral function is accessible to current state-of-the-art experiments with ultracold atoms. Unlike the dynamical structure factor, its measurement allows to identify  both Lieb-I and Lieb-II modes, since both have a diverging singularity in $A(k,\omega)$, respectively at $\epsilon_1(k)$ and at $-\epsilon_2(k)$.
The measurement of its  broad features and of the Lieb-II branch  will also  demonstrate the reach of a  beyond-mean-field regime. The knowledge of the spectral function is a key ingredient for the study of transport and out-of-equilibrium dynamics of strongly correlated bosons.

\begin{acknowledgments}
We acknowledges discussions with Roberta Citro, Aurelien Perrin and   H\'{e}l\`{e}ne Perrin.
NLG acknowledges financial support from the Academy of Finland Center of Excellence program (Project no. 312058), from the Turku Collegium for Science and Medicine (TCSM).
NLG acknowledges financial support from the COST action
"Quantum Technologies with Ultracold-Gases" (CA16221) and the Maupertieu programme for short-term travel grants. AM acknoledges funding from the SuperRing ANR project   (Grant  No.ANR-15-CE30-0012).
\end{acknowledgments}

\bibliography{bib}

\begin{thebibliography}{79}%
\makeatletter
\providecommand \@ifxundefined [1]{%
 \@ifx{#1\undefined}
}%
\providecommand \@ifnum [1]{%
 \ifnum #1\expandafter \@firstoftwo
 \else \expandafter \@secondoftwo
 \fi
}%
\providecommand \@ifx [1]{%
 \ifx #1\expandafter \@firstoftwo
 \else \expandafter \@secondoftwo
 \fi
}%
\providecommand \natexlab [1]{#1}%
\providecommand \enquote  [1]{``#1''}%
\providecommand \bibnamefont  [1]{#1}%
\providecommand \bibfnamefont [1]{#1}%
\providecommand \citenamefont [1]{#1}%
\providecommand \href@noop [0]{\@secondoftwo}%
\providecommand \href [0]{\begingroup \@sanitize@url \@href}%
\providecommand \@href[1]{\@@startlink{#1}\@@href}%
\providecommand \@@href[1]{\endgroup#1\@@endlink}%
\providecommand \@sanitize@url [0]{\catcode `\\12\catcode `\$12\catcode
  `\&12\catcode `\#12\catcode `\^12\catcode `\_12\catcode `\%12\relax}%
\providecommand \@@startlink[1]{}%
\providecommand \@@endlink[0]{}%
\providecommand \url  [0]{\begingroup\@sanitize@url \@url }%
\providecommand \@url [1]{\endgroup\@href {#1}{\urlprefix }}%
\providecommand \urlprefix  [0]{URL }%
\providecommand \Eprint [0]{\href }%
\providecommand \doibase [0]{http://dx.doi.org/}%
\providecommand \selectlanguage [0]{\@gobble}%
\providecommand \bibinfo  [0]{\@secondoftwo}%
\providecommand \bibfield  [0]{\@secondoftwo}%
\providecommand \translation [1]{[#1]}%
\providecommand \BibitemOpen [0]{}%
\providecommand \bibitemStop [0]{}%
\providecommand \bibitemNoStop [0]{.\EOS\space}%
\providecommand \EOS [0]{\spacefactor3000\relax}%
\providecommand \BibitemShut  [1]{\csname bibitem#1\endcsname}%
\let\auto@bib@innerbib\@empty
\bibitem [{\citenamefont {Girardeau}(1960)}]{Girardeau1960}%
  \BibitemOpen
  \bibfield  {author} {\bibinfo {author} {\bibfnamefont {M.}~\bibnamefont
  {Girardeau}},\ }\href {\doibase 10.1063/1.1703687} {\bibfield  {journal}
  {\bibinfo  {journal} {J. Math. Phys.}\ }\textbf {\bibinfo {volume} {1}},\
  \bibinfo {pages} {516} (\bibinfo {year} {1960})}\BibitemShut {NoStop}%
\bibitem [{\citenamefont {Bijl}(1937)}]{Bijl1937}%
  \BibitemOpen
  \bibfield  {author} {\bibinfo {author} {\bibfnamefont {A.}~\bibnamefont
  {Bijl}},\ }\href {\doibase 10.1016/S0031-8914(37)80057-9} {\bibfield
  {journal} {\bibinfo  {journal} {Physica}\ }\textbf {\bibinfo {volume} {4}},\
  \bibinfo {pages} {329} (\bibinfo {year} {1937})}\BibitemShut {NoStop}%
\bibitem [{\citenamefont {Olshanii}(1998)}]{Olshanii1998}%
  \BibitemOpen
  \bibfield  {author} {\bibinfo {author} {\bibfnamefont {M.}~\bibnamefont
  {Olshanii}},\ }\href {\doibase 10.1103/PhysRevLett.81.938} {\bibfield
  {journal} {\bibinfo  {journal} {Phys. Rev. Lett.}\ }\textbf {\bibinfo
  {volume} {81}},\ \bibinfo {pages} {938} (\bibinfo {year} {1998})}\BibitemShut
  {NoStop}%
\bibitem [{\citenamefont {Moritz}\ \emph {et~al.}(2003)\citenamefont {Moritz},
  \citenamefont {St{\"{o}}ferle}, \citenamefont {K{\"{o}}hl},\ and\
  \citenamefont {Esslinger}}]{Moritz2003}%
  \BibitemOpen
  \bibfield  {author} {\bibinfo {author} {\bibfnamefont {H.}~\bibnamefont
  {Moritz}}, \bibinfo {author} {\bibfnamefont {T.}~\bibnamefont
  {St{\"{o}}ferle}}, \bibinfo {author} {\bibfnamefont {M.}~\bibnamefont
  {K{\"{o}}hl}}, \ and\ \bibinfo {author} {\bibfnamefont {T.}~\bibnamefont
  {Esslinger}},\ }\href {\doibase 10.1103/PhysRevLett.91.250402} {\bibfield
  {journal} {\bibinfo  {journal} {Phys. Rev. Lett.}\ }\textbf {\bibinfo
  {volume} {91}},\ \bibinfo {pages} {250402} (\bibinfo {year}
  {2003})}\BibitemShut {NoStop}%
\bibitem [{\citenamefont {Kinoshita}\ \emph {et~al.}(2004)\citenamefont
  {Kinoshita}, \citenamefont {Wenger},\ and\ \citenamefont
  {Weiss}}]{Kinoshita2004}%
  \BibitemOpen
  \bibfield  {author} {\bibinfo {author} {\bibfnamefont {T.}~\bibnamefont
  {Kinoshita}}, \bibinfo {author} {\bibfnamefont {T.}~\bibnamefont {Wenger}}, \
  and\ \bibinfo {author} {\bibfnamefont {D.~S.}\ \bibnamefont {Weiss}},\ }\href
  {\doibase 10.1126/science.1100700} {\bibfield  {journal} {\bibinfo  {journal}
  {Science (80-. ).}\ }\textbf {\bibinfo {volume} {305}},\ \bibinfo {pages}
  {1125} (\bibinfo {year} {2004})}\BibitemShut {NoStop}%
\bibitem [{\citenamefont {Paredes}\ \emph {et~al.}(2004)\citenamefont
  {Paredes}, \citenamefont {Widera}, \citenamefont {Murg}, \citenamefont
  {Mandel}, \citenamefont {Foelling}, \citenamefont {Cirac}, \citenamefont
  {Shlyapnikov}, \citenamefont {Hansch},\ and\ \citenamefont
  {Bloch}}]{Paredes2004}%
  \BibitemOpen
  \bibfield  {author} {\bibinfo {author} {\bibfnamefont {B.}~\bibnamefont
  {Paredes}}, \bibinfo {author} {\bibfnamefont {a.}~\bibnamefont {Widera}},
  \bibinfo {author} {\bibfnamefont {V.}~\bibnamefont {Murg}}, \bibinfo {author}
  {\bibfnamefont {O.}~\bibnamefont {Mandel}}, \bibinfo {author} {\bibfnamefont
  {S.}~\bibnamefont {Foelling}}, \bibinfo {author} {\bibfnamefont
  {I.}~\bibnamefont {Cirac}}, \bibinfo {author} {\bibfnamefont {G.~V.}\
  \bibnamefont {Shlyapnikov}}, \bibinfo {author} {\bibfnamefont {T.~W.}\
  \bibnamefont {Hansch}}, \ and\ \bibinfo {author} {\bibfnamefont
  {I.}~\bibnamefont {Bloch}},\ }\href {\doibase 10.1038/nature02578.}
  {\bibfield  {journal} {\bibinfo  {journal} {Nature}\ }\textbf {\bibinfo
  {volume} {429}},\ \bibinfo {pages} {277} (\bibinfo {year}
  {2004})}\BibitemShut {NoStop}%
\bibitem [{\citenamefont {Kinoshita}\ \emph {et~al.}(2006)\citenamefont
  {Kinoshita}, \citenamefont {Wenger},\ and\ \citenamefont
  {Weiss}}]{Kinoshita2006}%
  \BibitemOpen
  \bibfield  {author} {\bibinfo {author} {\bibfnamefont {T.}~\bibnamefont
  {Kinoshita}}, \bibinfo {author} {\bibfnamefont {T.}~\bibnamefont {Wenger}}, \
  and\ \bibinfo {author} {\bibfnamefont {D.~S.}\ \bibnamefont {Weiss}},\ }\href
  {\doibase 10.1038/nature04693} {\bibfield  {journal} {\bibinfo  {journal}
  {Nature}\ }\textbf {\bibinfo {volume} {440}},\ \bibinfo {pages} {900}
  (\bibinfo {year} {2006})}\BibitemShut {NoStop}%
\bibitem [{\citenamefont {{Van Amerongen}}\ \emph {et~al.}(2008)\citenamefont
  {{Van Amerongen}}, \citenamefont {{Van Es}}, \citenamefont {Wicke},
  \citenamefont {Kheruntsyan},\ and\ \citenamefont {{Van
  Druten}}}]{VanAmerongen2008}%
  \BibitemOpen
  \bibfield  {author} {\bibinfo {author} {\bibfnamefont {A.~H.}\ \bibnamefont
  {{Van Amerongen}}}, \bibinfo {author} {\bibfnamefont {J.~J.}\ \bibnamefont
  {{Van Es}}}, \bibinfo {author} {\bibfnamefont {P.}~\bibnamefont {Wicke}},
  \bibinfo {author} {\bibfnamefont {K.~V.}\ \bibnamefont {Kheruntsyan}}, \ and\
  \bibinfo {author} {\bibfnamefont {N.~J.}\ \bibnamefont {{Van Druten}}},\
  }\href {\doibase 10.1103/PhysRevLett.100.090402} {\bibfield  {journal}
  {\bibinfo  {journal} {Phys. Rev. Lett.}\ }\textbf {\bibinfo {volume} {100}},\
  \bibinfo {pages} {13} (\bibinfo {year} {2008})}\BibitemShut {NoStop}%
\bibitem [{\citenamefont {Palzer}\ \emph {et~al.}(2009)\citenamefont {Palzer},
  \citenamefont {Zipkes}, \citenamefont {Sias},\ and\ \citenamefont
  {K{\"{o}}hl}}]{Palzer2009}%
  \BibitemOpen
  \bibfield  {author} {\bibinfo {author} {\bibfnamefont {S.}~\bibnamefont
  {Palzer}}, \bibinfo {author} {\bibfnamefont {C.}~\bibnamefont {Zipkes}},
  \bibinfo {author} {\bibfnamefont {C.}~\bibnamefont {Sias}}, \ and\ \bibinfo
  {author} {\bibfnamefont {M.}~\bibnamefont {K{\"{o}}hl}},\ }\href {\doibase
  10.1103/PhysRevLett.103.150601} {\bibfield  {journal} {\bibinfo  {journal}
  {Phys. Rev. Lett.}\ }\textbf {\bibinfo {volume} {103}},\ \bibinfo {pages} {1}
  (\bibinfo {year} {2009})}\BibitemShut {NoStop}%
\bibitem [{\citenamefont {Jacqmin}\ \emph {et~al.}(2011)\citenamefont
  {Jacqmin}, \citenamefont {Armijo}, \citenamefont {Berrada}, \citenamefont
  {Kheruntsyan},\ and\ \citenamefont {Bouchoule}}]{Jacqmin2011}%
  \BibitemOpen
  \bibfield  {author} {\bibinfo {author} {\bibfnamefont {T.}~\bibnamefont
  {Jacqmin}}, \bibinfo {author} {\bibfnamefont {J.}~\bibnamefont {Armijo}},
  \bibinfo {author} {\bibfnamefont {T.}~\bibnamefont {Berrada}}, \bibinfo
  {author} {\bibfnamefont {K.~V.}\ \bibnamefont {Kheruntsyan}}, \ and\ \bibinfo
  {author} {\bibfnamefont {I.}~\bibnamefont {Bouchoule}},\ }\href {\doibase
  10.1103/PhysRevLett.106.230405} {\bibfield  {journal} {\bibinfo  {journal}
  {Phys. Rev. Lett.}\ }\textbf {\bibinfo {volume} {106}},\ \bibinfo {pages} {1}
  (\bibinfo {year} {2011})}\BibitemShut {NoStop}%
\bibitem [{\citenamefont {Meinert}\ \emph {et~al.}(2017)\citenamefont
  {Meinert}, \citenamefont {Knap}, \citenamefont {Kirilov}, \citenamefont
  {Jag-Lauber}, \citenamefont {Zvonarev}, \citenamefont {Demler},\ and\
  \citenamefont {N{\"{a}}gerl}}]{Meinert2017}%
  \BibitemOpen
  \bibfield  {author} {\bibinfo {author} {\bibfnamefont {F.}~\bibnamefont
  {Meinert}}, \bibinfo {author} {\bibfnamefont {M.}~\bibnamefont {Knap}},
  \bibinfo {author} {\bibfnamefont {E.}~\bibnamefont {Kirilov}}, \bibinfo
  {author} {\bibfnamefont {K.}~\bibnamefont {Jag-Lauber}}, \bibinfo {author}
  {\bibfnamefont {M.~B.}\ \bibnamefont {Zvonarev}}, \bibinfo {author}
  {\bibfnamefont {E.}~\bibnamefont {Demler}}, \ and\ \bibinfo {author}
  {\bibfnamefont {H.~C.}\ \bibnamefont {N{\"{a}}gerl}},\ }\href {\doibase
  10.1126/science.aah6616} {\bibfield  {journal} {\bibinfo  {journal} {Science
  (80-. ).}\ }\textbf {\bibinfo {volume} {356}},\ \bibinfo {pages} {945}
  (\bibinfo {year} {2017})}\BibitemShut {NoStop}%
\bibitem [{\citenamefont {Wilson}\ \emph {et~al.}(2020)\citenamefont {Wilson},
  \citenamefont {Malvania}, \citenamefont {Le}, \citenamefont {Zhang},
  \citenamefont {Rigol},\ and\ \citenamefont {Weiss}}]{Wilson2020}%
  \BibitemOpen
  \bibfield  {author} {\bibinfo {author} {\bibfnamefont {J.~M.}\ \bibnamefont
  {Wilson}}, \bibinfo {author} {\bibfnamefont {N.}~\bibnamefont {Malvania}},
  \bibinfo {author} {\bibfnamefont {Y.}~\bibnamefont {Le}}, \bibinfo {author}
  {\bibfnamefont {Y.}~\bibnamefont {Zhang}}, \bibinfo {author} {\bibfnamefont
  {M.}~\bibnamefont {Rigol}}, \ and\ \bibinfo {author} {\bibfnamefont {D.~S.}\
  \bibnamefont {Weiss}},\ }\href {\doibase 10.1126/science.aaz0242} {\bibfield
  {journal} {\bibinfo  {journal} {Science}\ }\textbf {\bibinfo {volume}
  {367}},\ \bibinfo {pages} {1461} (\bibinfo {year} {2020})}\BibitemShut
  {NoStop}%
\bibitem [{\citenamefont {Lenard}(1966)}]{Lenard1966}%
  \BibitemOpen
  \bibfield  {author} {\bibinfo {author} {\bibfnamefont {A.}~\bibnamefont
  {Lenard}},\ }\href {\doibase 10.1063/1.1705029} {\bibfield  {journal}
  {\bibinfo  {journal} {J. Math. Phys.}\ }\textbf {\bibinfo {volume} {7}},\
  \bibinfo {pages} {1268} (\bibinfo {year} {1966})}\BibitemShut {NoStop}%
\bibitem [{\citenamefont {Vaidya}\ and\ \citenamefont
  {Tracy}(1979)}]{Vaidya1979}%
  \BibitemOpen
  \bibfield  {author} {\bibinfo {author} {\bibfnamefont {H.~G.}\ \bibnamefont
  {Vaidya}}\ and\ \bibinfo {author} {\bibfnamefont {C.~A.}\ \bibnamefont
  {Tracy}},\ }\href {\doibase 10.1103/PhysRevLett.42.3} {\bibfield  {journal}
  {\bibinfo  {journal} {Phys. Rev. Lett.}\ }\textbf {\bibinfo {volume} {42}},\
  \bibinfo {pages} {3} (\bibinfo {year} {1979})}\BibitemShut {NoStop}%
\bibitem [{\citenamefont {Jimbo}\ \emph {et~al.}(1980)\citenamefont {Jimbo},
  \citenamefont {Miwa}, \citenamefont {M{\^{o}}ri},\ and\ \citenamefont
  {Sato}}]{Jimbo1980}%
  \BibitemOpen
  \bibfield  {author} {\bibinfo {author} {\bibfnamefont {M.}~\bibnamefont
  {Jimbo}}, \bibinfo {author} {\bibfnamefont {T.}~\bibnamefont {Miwa}},
  \bibinfo {author} {\bibfnamefont {Y.}~\bibnamefont {M{\^{o}}ri}}, \ and\
  \bibinfo {author} {\bibfnamefont {M.}~\bibnamefont {Sato}},\ }\href {\doibase
  10.1016/0167-2789(80)90006-8} {\bibfield  {journal} {\bibinfo  {journal}
  {Phys. D Nonlinear Phenom.}\ }\textbf {\bibinfo {volume} {1}},\ \bibinfo
  {pages} {80} (\bibinfo {year} {1980})}\BibitemShut {NoStop}%
\bibitem [{\citenamefont {Girardeau}\ \emph {et~al.}(2001)\citenamefont
  {Girardeau}, \citenamefont {Wright},\ and\ \citenamefont
  {Triscari}}]{Girardeau2001}%
  \BibitemOpen
  \bibfield  {author} {\bibinfo {author} {\bibfnamefont {M.~D.}\ \bibnamefont
  {Girardeau}}, \bibinfo {author} {\bibfnamefont {E.~M.}\ \bibnamefont
  {Wright}}, \ and\ \bibinfo {author} {\bibfnamefont {J.~M.}\ \bibnamefont
  {Triscari}},\ }\href {\doibase 10.1103/PhysRevA.63.033601} {\bibfield
  {journal} {\bibinfo  {journal} {Phys. Rev. A - At. Mol. Opt. Phys.}\ }\textbf
  {\bibinfo {volume} {63}},\ \bibinfo {pages} {1} (\bibinfo {year}
  {2001})}\BibitemShut {NoStop}%
\bibitem [{\citenamefont {Papenbrock}(2003)}]{Papenbrock2003}%
  \BibitemOpen
  \bibfield  {author} {\bibinfo {author} {\bibfnamefont {T.}~\bibnamefont
  {Papenbrock}},\ }\href {\doibase 10.1103/PhysRevA.67.041601} {\bibfield
  {journal} {\bibinfo  {journal} {Phys. Rev. A - At. Mol. Opt. Phys.}\ }\textbf
  {\bibinfo {volume} {67}},\ \bibinfo {pages} {4} (\bibinfo {year}
  {2003})}\BibitemShut {NoStop}%
\bibitem [{\citenamefont {Forrester}\ \emph
  {et~al.}(2003{\natexlab{a}})\citenamefont {Forrester}, \citenamefont
  {Frankel}, \citenamefont {Garoni},\ and\ \citenamefont
  {Witte}}]{Forrester2002}%
  \BibitemOpen
  \bibfield  {author} {\bibinfo {author} {\bibfnamefont {P.}~\bibnamefont
  {Forrester}}, \bibinfo {author} {\bibfnamefont {N.}~\bibnamefont {Frankel}},
  \bibinfo {author} {\bibfnamefont {T.}~\bibnamefont {Garoni}}, \ and\ \bibinfo
  {author} {\bibfnamefont {N.}~\bibnamefont {Witte}},\ }\href {\doibase
  10.1007/s00220-003-0851-3} {\bibfield  {journal} {\bibinfo  {journal}
  {Commun. Math. Phys.}\ }\textbf {\bibinfo {volume} {238}},\ \bibinfo {pages}
  {257} (\bibinfo {year} {2003}{\natexlab{a}})}\BibitemShut {NoStop}%
\bibitem [{\citenamefont {Forrester}\ \emph
  {et~al.}(2003{\natexlab{b}})\citenamefont {Forrester}, \citenamefont
  {Frankel}, \citenamefont {Garoni},\ and\ \citenamefont
  {Witte}}]{Forrester2003}%
  \BibitemOpen
  \bibfield  {author} {\bibinfo {author} {\bibfnamefont {P.~J.}\ \bibnamefont
  {Forrester}}, \bibinfo {author} {\bibfnamefont {N.~E.}\ \bibnamefont
  {Frankel}}, \bibinfo {author} {\bibfnamefont {T.~M.}\ \bibnamefont {Garoni}},
  \ and\ \bibinfo {author} {\bibfnamefont {N.~S.}\ \bibnamefont {Witte}},\
  }\href {\doibase 10.1103/PhysRevA.67.043607} {\bibfield  {journal} {\bibinfo
  {journal} {Phys. Rev. A - At. Mol. Opt. Phys.}\ }\textbf {\bibinfo {volume}
  {67}},\ \bibinfo {pages} {17} (\bibinfo {year}
  {2003}{\natexlab{b}})}\BibitemShut {NoStop}%
\bibitem [{\citenamefont {Castin}(2004)}]{Castin2004}%
  \BibitemOpen
  \bibfield  {author} {\bibinfo {author} {\bibfnamefont {Y.}~\bibnamefont
  {Castin}},\ }\href {\doibase 10.1051/jp4:2004116004} {\bibfield  {journal}
  {\bibinfo  {journal} {J. Phys. IV Fr.}\ }\textbf {\bibinfo {volume} {116}},\
  \bibinfo {pages} {1} (\bibinfo {year} {2004})}\BibitemShut {NoStop}%
\bibitem [{\citenamefont {Rigol}(2005)}]{Rigol2005}%
  \BibitemOpen
  \bibfield  {author} {\bibinfo {author} {\bibfnamefont {M.}~\bibnamefont
  {Rigol}},\ }\href {\doibase 10.1103/PhysRevA.72.063607} {\bibfield  {journal}
  {\bibinfo  {journal} {Phys. Rev. A - At. Mol. Opt. Phys.}\ }\textbf {\bibinfo
  {volume} {72}},\ \bibinfo {pages} {1} (\bibinfo {year} {2005})}\BibitemShut
  {NoStop}%
\bibitem [{\citenamefont {Yukalov}\ and\ \citenamefont
  {Girardeau}(2005)}]{Yukalov2005}%
  \BibitemOpen
  \bibfield  {author} {\bibinfo {author} {\bibfnamefont {V.~I.}\ \bibnamefont
  {Yukalov}}\ and\ \bibinfo {author} {\bibfnamefont {M.~D.}\ \bibnamefont
  {Girardeau}},\ }\href {\doibase 10.1002/lapl.200510011} {\bibfield  {journal}
  {\bibinfo  {journal} {Laser Phys. Lett.}\ }\textbf {\bibinfo {volume} {2}},\
  \bibinfo {pages} {375} (\bibinfo {year} {2005})}\BibitemShut {NoStop}%
\bibitem [{\citenamefont {Vignolo}\ and\ \citenamefont
  {Minguzzi}(2013)}]{Vignolo2013}%
  \BibitemOpen
  \bibfield  {author} {\bibinfo {author} {\bibfnamefont {P.}~\bibnamefont
  {Vignolo}}\ and\ \bibinfo {author} {\bibfnamefont {A.}~\bibnamefont
  {Minguzzi}},\ }\href {\doibase 10.1103/PhysRevLett.110.020403} {\bibfield
  {journal} {\bibinfo  {journal} {Phys. Rev. Lett.}\ }\textbf {\bibinfo
  {volume} {110}},\ \bibinfo {pages} {1} (\bibinfo {year} {2013})}\BibitemShut
  {NoStop}%
\bibitem [{\citenamefont {Garc{\'{i}}a-March}\ \emph
  {et~al.}(2015)\citenamefont {Garc{\'{i}}a-March}, \citenamefont {Yuste},
  \citenamefont {Juli{\'{a}}-D{\'{i}}az},\ and\ \citenamefont
  {Polls}}]{Garcia-March2015}%
  \BibitemOpen
  \bibfield  {author} {\bibinfo {author} {\bibfnamefont {M.~A.}\ \bibnamefont
  {Garc{\'{i}}a-March}}, \bibinfo {author} {\bibfnamefont {A.}~\bibnamefont
  {Yuste}}, \bibinfo {author} {\bibfnamefont {B.}~\bibnamefont
  {Juli{\'{a}}-D{\'{i}}az}}, \ and\ \bibinfo {author} {\bibfnamefont
  {A.}~\bibnamefont {Polls}},\ }\href {\doibase 10.1103/PhysRevA.92.033621}
  {\bibfield  {journal} {\bibinfo  {journal} {Phys. Rev. A}\ }\textbf {\bibinfo
  {volume} {92}},\ \bibinfo {pages} {033621} (\bibinfo {year}
  {2015})}\BibitemShut {NoStop}%
\bibitem [{\citenamefont {Xu}\ and\ \citenamefont {Rigol}(2017)}]{Xu2017}%
  \BibitemOpen
  \bibfield  {author} {\bibinfo {author} {\bibfnamefont {W.}~\bibnamefont
  {Xu}}\ and\ \bibinfo {author} {\bibfnamefont {M.}~\bibnamefont {Rigol}},\
  }\href {\doibase 10.1103/PhysRevA.95.033617} {\bibfield  {journal} {\bibinfo
  {journal} {Phys. Rev. A}\ }\textbf {\bibinfo {volume} {95}},\ \bibinfo
  {pages} {1} (\bibinfo {year} {2017})}\BibitemShut {NoStop}%
\bibitem [{\citenamefont {Settino}\ \emph {et~al.}(2017)\citenamefont
  {Settino}, \citenamefont {{Lo Gullo}}, \citenamefont {Sindona}, \citenamefont
  {Goold},\ and\ \citenamefont {Plastina}}]{Settino2017}%
  \BibitemOpen
  \bibfield  {author} {\bibinfo {author} {\bibfnamefont {J.}~\bibnamefont
  {Settino}}, \bibinfo {author} {\bibfnamefont {N.}~\bibnamefont {{Lo Gullo}}},
  \bibinfo {author} {\bibfnamefont {A.}~\bibnamefont {Sindona}}, \bibinfo
  {author} {\bibfnamefont {J.}~\bibnamefont {Goold}}, \ and\ \bibinfo {author}
  {\bibfnamefont {F.}~\bibnamefont {Plastina}},\ }\href {\doibase
  10.1103/PhysRevA.95.033605} {\bibfield  {journal} {\bibinfo  {journal} {Phys.
  Rev. A}\ }\textbf {\bibinfo {volume} {95}},\ \bibinfo {pages} {1} (\bibinfo
  {year} {2017})}\BibitemShut {NoStop}%
\bibitem [{\citenamefont {Atas}\ \emph {et~al.}(2017)\citenamefont {Atas},
  \citenamefont {Gangardt}, \citenamefont {Bouchoule},\ and\ \citenamefont
  {Kheruntsyan}}]{Atas2017}%
  \BibitemOpen
  \bibfield  {author} {\bibinfo {author} {\bibfnamefont {Y.~Y.}\ \bibnamefont
  {Atas}}, \bibinfo {author} {\bibfnamefont {D.~M.}\ \bibnamefont {Gangardt}},
  \bibinfo {author} {\bibfnamefont {I.}~\bibnamefont {Bouchoule}}, \ and\
  \bibinfo {author} {\bibfnamefont {K.~V.}\ \bibnamefont {Kheruntsyan}},\
  }\href {\doibase 10.1103/PhysRevA.95.043622} {\bibfield  {journal} {\bibinfo
  {journal} {Phys. Rev. A}\ }\textbf {\bibinfo {volume} {95}},\ \bibinfo
  {pages} {1} (\bibinfo {year} {2017})}\BibitemShut {NoStop}%
\bibitem [{\citenamefont {Colcelli}\ \emph {et~al.}(2018)\citenamefont
  {Colcelli}, \citenamefont {Viti}, \citenamefont {Mussardo},\ and\
  \citenamefont {Trombettoni}}]{Colcelli2018}%
  \BibitemOpen
  \bibfield  {author} {\bibinfo {author} {\bibfnamefont {A.}~\bibnamefont
  {Colcelli}}, \bibinfo {author} {\bibfnamefont {J.}~\bibnamefont {Viti}},
  \bibinfo {author} {\bibfnamefont {G.}~\bibnamefont {Mussardo}}, \ and\
  \bibinfo {author} {\bibfnamefont {A.}~\bibnamefont {Trombettoni}},\ }\href
  {\doibase 10.1103/PhysRevA.98.063633} {\bibfield  {journal} {\bibinfo
  {journal} {Phys. Rev. A}\ }\textbf {\bibinfo {volume} {98}},\ \bibinfo
  {pages} {063633} (\bibinfo {year} {2018})}\BibitemShut {NoStop}%
\bibitem [{\citenamefont {Lang}(2018)}]{Lang2017}%
  \BibitemOpen
  \bibfield  {author} {\bibinfo {author} {\bibfnamefont {G.}~\bibnamefont
  {Lang}},\ }\href {\doibase 10.1007/978-3-030-05285-0} {\emph {\bibinfo
  {title} {{Correlations in Low-Dimensional Quantum Gases}}}},\ Springer
  Theses\ (\bibinfo  {publisher} {Springer International Publishing},\ \bibinfo
  {address} {Cham},\ \bibinfo {year} {2018})\BibitemShut {NoStop}%
\bibitem [{\citenamefont {Brun}\ and\ \citenamefont {Dubail}(2017)}]{Brun2017}%
  \BibitemOpen
  \bibfield  {author} {\bibinfo {author} {\bibfnamefont {Y.}~\bibnamefont
  {Brun}}\ and\ \bibinfo {author} {\bibfnamefont {J.}~\bibnamefont {Dubail}},\
  }\href {\doibase 10.21468/SciPostPhys.2.2.012} {\bibfield  {journal}
  {\bibinfo  {journal} {SciPost Phys.}\ }\textbf {\bibinfo {volume} {2}},\
  \bibinfo {pages} {1} (\bibinfo {year} {2017})}\BibitemShut {NoStop}%
\bibitem [{\citenamefont {Lenard}(1964)}]{Lenard1964}%
  \BibitemOpen
  \bibfield  {author} {\bibinfo {author} {\bibfnamefont {A.}~\bibnamefont
  {Lenard}},\ }\href {\doibase 10.1063/1.1704196} {\bibfield  {journal}
  {\bibinfo  {journal} {J. Math. Phys.}\ }\textbf {\bibinfo {volume} {5}},\
  \bibinfo {pages} {930} (\bibinfo {year} {1964})}\BibitemShut {NoStop}%
\bibitem [{\citenamefont {Minguzzi}\ \emph {et~al.}(2002)\citenamefont
  {Minguzzi}, \citenamefont {Vignolo},\ and\ \citenamefont
  {Tosi}}]{Minguzzi2002}%
  \BibitemOpen
  \bibfield  {author} {\bibinfo {author} {\bibfnamefont {A.}~\bibnamefont
  {Minguzzi}}, \bibinfo {author} {\bibfnamefont {P.}~\bibnamefont {Vignolo}}, \
  and\ \bibinfo {author} {\bibfnamefont {M.~P.}\ \bibnamefont {Tosi}},\ }\href
  {\doibase 10.1016/S0375-9601(02)00042-7} {\bibfield  {journal} {\bibinfo
  {journal} {Phys. Lett. Sect. A Gen. At. Solid State Phys.}\ }\textbf
  {\bibinfo {volume} {294}},\ \bibinfo {pages} {222} (\bibinfo {year}
  {2002})}\BibitemShut {NoStop}%
\bibitem [{\citenamefont {Lapeyre}\ \emph {et~al.}(2002)\citenamefont
  {Lapeyre}, \citenamefont {Girardeau},\ and\ \citenamefont
  {Wright}}]{Lapeyre2002}%
  \BibitemOpen
  \bibfield  {author} {\bibinfo {author} {\bibfnamefont {G.~J.}\ \bibnamefont
  {Lapeyre}}, \bibinfo {author} {\bibfnamefont {M.~D.}\ \bibnamefont
  {Girardeau}}, \ and\ \bibinfo {author} {\bibfnamefont {E.~M.}\ \bibnamefont
  {Wright}},\ }\href {\doibase 10.1103/PhysRevA.66.023606} {\bibfield
  {journal} {\bibinfo  {journal} {Phys. Rev. A - At. Mol. Opt. Phys.}\ }\textbf
  {\bibinfo {volume} {66}},\ \bibinfo {pages} {1} (\bibinfo {year}
  {2002})}\BibitemShut {NoStop}%
\bibitem [{\citenamefont {Rigol}\ and\ \citenamefont
  {Muramatsu}(2004)}]{Rigol2004}%
  \BibitemOpen
  \bibfield  {author} {\bibinfo {author} {\bibfnamefont {M.}~\bibnamefont
  {Rigol}}\ and\ \bibinfo {author} {\bibfnamefont {A.}~\bibnamefont
  {Muramatsu}},\ }\href {\doibase 10.1103/PhysRevLett.94.240403} {\bibfield
  {journal} {\bibinfo  {journal} {Opt.{\~{}} Commun.{\~{}}}\ }\textbf {\bibinfo
  {volume} {243}},\ \bibinfo {pages} {33} (\bibinfo {year} {2004})}\BibitemShut
  {NoStop}%
\bibitem [{\citenamefont {Minguzzi}\ and\ \citenamefont
  {Gangardt}(2005)}]{Minguzzi2005}%
  \BibitemOpen
  \bibfield  {author} {\bibinfo {author} {\bibfnamefont {A.}~\bibnamefont
  {Minguzzi}}\ and\ \bibinfo {author} {\bibfnamefont {D.~M.}\ \bibnamefont
  {Gangardt}},\ }\href {\doibase 10.1103/PhysRevLett.94.240404} {\bibfield
  {journal} {\bibinfo  {journal} {Phys. Rev. Lett.}\ }\textbf {\bibinfo
  {volume} {94}},\ \bibinfo {pages} {240404} (\bibinfo {year}
  {2005})}\BibitemShut {NoStop}%
\bibitem [{\citenamefont {Rigol}\ and\ \citenamefont
  {Muramatsu}(2006)}]{Rigol2006}%
  \BibitemOpen
  \bibfield  {author} {\bibinfo {author} {\bibfnamefont {M.}~\bibnamefont
  {Rigol}}\ and\ \bibinfo {author} {\bibfnamefont {A.}~\bibnamefont
  {Muramatsu}},\ }\href {\doibase 10.1134/S1054660X06020253} {\bibfield
  {journal} {\bibinfo  {journal} {Laser Phys.}\ }\textbf {\bibinfo {volume}
  {16}},\ \bibinfo {pages} {348} (\bibinfo {year} {2006})}\BibitemShut
  {NoStop}%
\bibitem [{\citenamefont {Pezer}\ and\ \citenamefont
  {Buljan}(2007)}]{Pezer2007}%
  \BibitemOpen
  \bibfield  {author} {\bibinfo {author} {\bibfnamefont {R.}~\bibnamefont
  {Pezer}}\ and\ \bibinfo {author} {\bibfnamefont {H.}~\bibnamefont {Buljan}},\
  }\href {\doibase 10.1103/PhysRevLett.98.240403} {\bibfield  {journal}
  {\bibinfo  {journal} {Phys. Rev. Lett.}\ }\textbf {\bibinfo {volume} {98}},\
  \bibinfo {pages} {240403} (\bibinfo {year} {2007})}\BibitemShut {NoStop}%
\bibitem [{\citenamefont {Deng}\ \emph {et~al.}(2008)\citenamefont {Deng},
  \citenamefont {Citro}, \citenamefont {Minguzzi},\ and\ \citenamefont
  {Orignac}}]{Deng2008}%
  \BibitemOpen
  \bibfield  {author} {\bibinfo {author} {\bibfnamefont {X.}~\bibnamefont
  {Deng}}, \bibinfo {author} {\bibfnamefont {R.}~\bibnamefont {Citro}},
  \bibinfo {author} {\bibfnamefont {A.}~\bibnamefont {Minguzzi}}, \ and\
  \bibinfo {author} {\bibfnamefont {E.}~\bibnamefont {Orignac}},\ }\href
  {\doibase 10.1103/PhysRevA.78.013625} {\bibfield  {journal} {\bibinfo
  {journal} {Phys. Rev. A - At. Mol. Opt. Phys.}\ }\textbf {\bibinfo {volume}
  {78}},\ \bibinfo {pages} {013625} (\bibinfo {year} {2008})}\BibitemShut
  {NoStop}%
\bibitem [{\citenamefont {Das}\ \emph {et~al.}(2002)\citenamefont {Das},
  \citenamefont {Girardeau},\ and\ \citenamefont {Wright}}]{Das2002}%
  \BibitemOpen
  \bibfield  {author} {\bibinfo {author} {\bibfnamefont {K.~K.}\ \bibnamefont
  {Das}}, \bibinfo {author} {\bibfnamefont {M.~D.}\ \bibnamefont {Girardeau}},
  \ and\ \bibinfo {author} {\bibfnamefont {E.~M.}\ \bibnamefont {Wright}},\
  }\href {\doibase 10.1103/PhysRevLett.89.170404} {\bibfield  {journal}
  {\bibinfo  {journal} {Phys. Rev. Lett.}\ }\textbf {\bibinfo {volume} {89}},\
  \bibinfo {pages} {17} (\bibinfo {year} {2002})}\BibitemShut {NoStop}%
\bibitem [{\citenamefont {Berman}\ \emph {et~al.}(2004)\citenamefont {Berman},
  \citenamefont {Borgonovi}, \citenamefont {Izrailev},\ and\ \citenamefont
  {Smerzi}}]{Berman2004}%
  \BibitemOpen
  \bibfield  {author} {\bibinfo {author} {\bibfnamefont {G.~P.}\ \bibnamefont
  {Berman}}, \bibinfo {author} {\bibfnamefont {F.}~\bibnamefont {Borgonovi}},
  \bibinfo {author} {\bibfnamefont {F.~M.}\ \bibnamefont {Izrailev}}, \ and\
  \bibinfo {author} {\bibfnamefont {A.}~\bibnamefont {Smerzi}},\ }\href
  {\doibase 10.1103/PhysRevLett.92.030404} {\bibfield  {journal} {\bibinfo
  {journal} {Phys. Rev. Lett.}\ }\textbf {\bibinfo {volume} {92}},\ \bibinfo
  {pages} {4} (\bibinfo {year} {2004})}\BibitemShut {NoStop}%
\bibitem [{\citenamefont {Rigol}\ \emph {et~al.}(2007)\citenamefont {Rigol},
  \citenamefont {Dunjko}, \citenamefont {Yurovsky},\ and\ \citenamefont
  {Olshanii}}]{Rigol2007}%
  \BibitemOpen
  \bibfield  {author} {\bibinfo {author} {\bibfnamefont {M.}~\bibnamefont
  {Rigol}}, \bibinfo {author} {\bibfnamefont {V.}~\bibnamefont {Dunjko}},
  \bibinfo {author} {\bibfnamefont {V.}~\bibnamefont {Yurovsky}}, \ and\
  \bibinfo {author} {\bibfnamefont {M.}~\bibnamefont {Olshanii}},\ }\href
  {\doibase 10.1103/PhysRevLett.98.050405} {\bibfield  {journal} {\bibinfo
  {journal} {Phys. Rev. Lett.}\ }\textbf {\bibinfo {volume} {98}},\ \bibinfo
  {pages} {050405} (\bibinfo {year} {2007})}\BibitemShut {NoStop}%
\bibitem [{\citenamefont {Kormos}\ \emph {et~al.}(2014)\citenamefont {Kormos},
  \citenamefont {Collura},\ and\ \citenamefont {Calabrese}}]{Kormos2014}%
  \BibitemOpen
  \bibfield  {author} {\bibinfo {author} {\bibfnamefont {M.}~\bibnamefont
  {Kormos}}, \bibinfo {author} {\bibfnamefont {M.}~\bibnamefont {Collura}}, \
  and\ \bibinfo {author} {\bibfnamefont {P.}~\bibnamefont {Calabrese}},\ }\href
  {\doibase 10.1103/PhysRevA.89.013609} {\bibfield  {journal} {\bibinfo
  {journal} {Phys. Rev. A - At. Mol. Opt. Phys.}\ }\textbf {\bibinfo {volume}
  {89}} (\bibinfo {year} {2014}),\ 10.1103/PhysRevA.89.013609}\BibitemShut
  {NoStop}%
\bibitem [{\citenamefont {Boumaza}\ and\ \citenamefont
  {Bencheikh}(2017)}]{Boumaza2017}%
  \BibitemOpen
  \bibfield  {author} {\bibinfo {author} {\bibfnamefont {R.}~\bibnamefont
  {Boumaza}}\ and\ \bibinfo {author} {\bibfnamefont {K.}~\bibnamefont
  {Bencheikh}},\ }\href {\doibase 10.1088/1751-8121/aa9363} {\bibfield
  {journal} {\bibinfo  {journal} {J. Phys. A Math. Theor.}\ }\textbf {\bibinfo
  {volume} {50}} (\bibinfo {year} {2017}),\
  10.1088/1751-8121/aa9363}\BibitemShut {NoStop}%
\bibitem [{\citenamefont {Bastianello}\ \emph {et~al.}(2017)\citenamefont
  {Bastianello}, \citenamefont {Collura},\ and\ \citenamefont
  {Sotiriadis}}]{Bastianello2017}%
  \BibitemOpen
  \bibfield  {author} {\bibinfo {author} {\bibfnamefont {A.}~\bibnamefont
  {Bastianello}}, \bibinfo {author} {\bibfnamefont {M.}~\bibnamefont
  {Collura}}, \ and\ \bibinfo {author} {\bibfnamefont {S.}~\bibnamefont
  {Sotiriadis}},\ }\href {\doibase 10.1103/PhysRevB.95.174303} {\bibfield
  {journal} {\bibinfo  {journal} {Phys. Rev. B}\ }\textbf {\bibinfo {volume}
  {95}},\ \bibinfo {pages} {1} (\bibinfo {year} {2017})}\BibitemShut {NoStop}%
\bibitem [{\citenamefont {{Yago Malo}}\ \emph {et~al.}(2018)\citenamefont
  {{Yago Malo}}, \citenamefont {{Van Nieuwenburg}}, \citenamefont {Fischer},\
  and\ \citenamefont {Daley}}]{YagoMalo2018}%
  \BibitemOpen
  \bibfield  {author} {\bibinfo {author} {\bibfnamefont {J.}~\bibnamefont
  {{Yago Malo}}}, \bibinfo {author} {\bibfnamefont {E.~P.}\ \bibnamefont {{Van
  Nieuwenburg}}}, \bibinfo {author} {\bibfnamefont {M.~H.}\ \bibnamefont
  {Fischer}}, \ and\ \bibinfo {author} {\bibfnamefont {A.~J.}\ \bibnamefont
  {Daley}},\ }\href {\doibase 10.1103/PhysRevA.97.053614} {\bibfield  {journal}
  {\bibinfo  {journal} {Phys. Rev. A}\ }\textbf {\bibinfo {volume} {97}},\
  \bibinfo {pages} {1} (\bibinfo {year} {2018})}\BibitemShut {NoStop}%
\bibitem [{\citenamefont {Mikkelsen}\ \emph {et~al.}(2018)\citenamefont
  {Mikkelsen}, \citenamefont {Fogarty},\ and\ \citenamefont
  {Busch}}]{Mikkelsen2018}%
  \BibitemOpen
  \bibfield  {author} {\bibinfo {author} {\bibfnamefont {M.}~\bibnamefont
  {Mikkelsen}}, \bibinfo {author} {\bibfnamefont {T.}~\bibnamefont {Fogarty}},
  \ and\ \bibinfo {author} {\bibfnamefont {T.}~\bibnamefont {Busch}},\ }\href
  {\doibase 10.1088/1367-2630/aae98e} {\bibfield  {journal} {\bibinfo
  {journal} {New Journal of Physics}\ }\textbf {\bibinfo {volume} {20}},\
  \bibinfo {pages} {113011} (\bibinfo {year} {2018})}\BibitemShut {NoStop}%
\bibitem [{\citenamefont {Damascelli}(2004)}]{Damascelli2004}%
  \BibitemOpen
  \bibfield  {author} {\bibinfo {author} {\bibfnamefont {A.}~\bibnamefont
  {Damascelli}},\ }\href {\doibase 10.1238/Physica.Topical.109a00061}
  {\bibfield  {journal} {\bibinfo  {journal} {Phys. Scr.}\ }\textbf {\bibinfo
  {volume} {T109}},\ \bibinfo {pages} {61} (\bibinfo {year}
  {2004})}\BibitemShut {NoStop}%
\bibitem [{\citenamefont {Stewart}\ \emph {et~al.}(2008)\citenamefont
  {Stewart}, \citenamefont {Gaebler},\ and\ \citenamefont {Jin}}]{Stewart2008}%
  \BibitemOpen
  \bibfield  {author} {\bibinfo {author} {\bibfnamefont {J.~T.}\ \bibnamefont
  {Stewart}}, \bibinfo {author} {\bibfnamefont {J.~P.}\ \bibnamefont
  {Gaebler}}, \ and\ \bibinfo {author} {\bibfnamefont {D.~S.}\ \bibnamefont
  {Jin}},\ }\href {\doibase 10.1038/nature07172} {\bibfield  {journal}
  {\bibinfo  {journal} {Nature}\ }\textbf {\bibinfo {volume} {454}},\ \bibinfo
  {pages} {744} (\bibinfo {year} {2008})}\BibitemShut {NoStop}%
\bibitem [{\citenamefont {Dao}\ \emph {et~al.}(2009)\citenamefont {Dao},
  \citenamefont {Carusotto},\ and\ \citenamefont {Georges}}]{Dao2009}%
  \BibitemOpen
  \bibfield  {author} {\bibinfo {author} {\bibfnamefont {T.~L.}\ \bibnamefont
  {Dao}}, \bibinfo {author} {\bibfnamefont {I.}~\bibnamefont {Carusotto}}, \
  and\ \bibinfo {author} {\bibfnamefont {A.}~\bibnamefont {Georges}},\ }\href
  {\doibase 10.1103/PhysRevA.80.023627} {\bibfield  {journal} {\bibinfo
  {journal} {Phys. Rev. A - At. Mol. Opt. Phys.}\ }\textbf {\bibinfo {volume}
  {80}},\ \bibinfo {pages} {1} (\bibinfo {year} {2009})}\BibitemShut {NoStop}%
\bibitem [{\citenamefont {Volchkov}\ \emph {et~al.}(2018)\citenamefont
  {Volchkov}, \citenamefont {Pasek}, \citenamefont {Denechaud}, \citenamefont
  {Mukhtar}, \citenamefont {Aspect}, \citenamefont {Delande},\ and\
  \citenamefont {Josse}}]{Volchkov2018}%
  \BibitemOpen
  \bibfield  {author} {\bibinfo {author} {\bibfnamefont {V.~V.}\ \bibnamefont
  {Volchkov}}, \bibinfo {author} {\bibfnamefont {M.}~\bibnamefont {Pasek}},
  \bibinfo {author} {\bibfnamefont {V.}~\bibnamefont {Denechaud}}, \bibinfo
  {author} {\bibfnamefont {M.}~\bibnamefont {Mukhtar}}, \bibinfo {author}
  {\bibfnamefont {A.}~\bibnamefont {Aspect}}, \bibinfo {author} {\bibfnamefont
  {D.}~\bibnamefont {Delande}}, \ and\ \bibinfo {author} {\bibfnamefont
  {V.}~\bibnamefont {Josse}},\ }\href {\doibase 10.1103/PhysRevLett.120.060404}
  {\bibfield  {journal} {\bibinfo  {journal} {Phys. Rev. Lett.}\ }\textbf
  {\bibinfo {volume} {120}} (\bibinfo {year} {2018}),\
  10.1103/PhysRevLett.120.060404}\BibitemShut {NoStop}%
\bibitem [{\citenamefont {Bohrdt}\ \emph {et~al.}(2018)\citenamefont {Bohrdt},
  \citenamefont {Greif}, \citenamefont {Demler}, \citenamefont {Knap},\ and\
  \citenamefont {Grusdt}}]{Bohrdt2018}%
  \BibitemOpen
  \bibfield  {author} {\bibinfo {author} {\bibfnamefont {A.}~\bibnamefont
  {Bohrdt}}, \bibinfo {author} {\bibfnamefont {D.}~\bibnamefont {Greif}},
  \bibinfo {author} {\bibfnamefont {E.}~\bibnamefont {Demler}}, \bibinfo
  {author} {\bibfnamefont {M.}~\bibnamefont {Knap}}, \ and\ \bibinfo {author}
  {\bibfnamefont {F.}~\bibnamefont {Grusdt}},\ }\href {\doibase
  10.1103/PhysRevB.97.125117} {\bibfield  {journal} {\bibinfo  {journal} {Phys.
  Rev. B}\ }\textbf {\bibinfo {volume} {97}},\ \bibinfo {pages} {125117}
  (\bibinfo {year} {2018})}\BibitemShut {NoStop}%
\bibitem [{\citenamefont {Stefanucci}\ and\ \citenamefont {{Van
  Leeuwen}}(2010)}]{Stefanucci2010}%
  \BibitemOpen
  \bibfield  {author} {\bibinfo {author} {\bibfnamefont {G.}~\bibnamefont
  {Stefanucci}}\ and\ \bibinfo {author} {\bibfnamefont {R.}~\bibnamefont {{Van
  Leeuwen}}},\ }\href {\doibase 10.1017/CBO9781139023979} {\emph {\bibinfo
  {title} {Nonequilibrium Many-Body Theory Quantum Syst. A Mod. Introd.}}},\
  Vol.\ \bibinfo {volume} {9780521766}\ (\bibinfo  {publisher} {Cambridge
  University Press},\ \bibinfo {year} {2010})\ pp.\ \bibinfo {pages}
  {1--600}\BibitemShut {NoStop}%
\bibitem [{\citenamefont {Tuovinen}\ \emph {et~al.}(2014)\citenamefont
  {Tuovinen}, \citenamefont {Perfetto}, \citenamefont {Stefanucci},\ and\
  \citenamefont {van Leeuwen}}]{Tuovinen2014}%
  \BibitemOpen
  \bibfield  {author} {\bibinfo {author} {\bibfnamefont {R.}~\bibnamefont
  {Tuovinen}}, \bibinfo {author} {\bibfnamefont {E.}~\bibnamefont {Perfetto}},
  \bibinfo {author} {\bibfnamefont {G.}~\bibnamefont {Stefanucci}}, \ and\
  \bibinfo {author} {\bibfnamefont {R.}~\bibnamefont {van Leeuwen}},\ }\href
  {\doibase 10.1103/PhysRevB.89.085131} {\bibfield  {journal} {\bibinfo
  {journal} {Phys. Rev. B}\ }\textbf {\bibinfo {volume} {89}},\ \bibinfo
  {pages} {085131} (\bibinfo {year} {2014})}\BibitemShut {NoStop}%
\bibitem [{\citenamefont {Ridley}\ and\ \citenamefont
  {Tuovinen}(2017)}]{Tuovinen2017}%
  \BibitemOpen
  \bibfield  {author} {\bibinfo {author} {\bibfnamefont {M.}~\bibnamefont
  {Ridley}}\ and\ \bibinfo {author} {\bibfnamefont {R.}~\bibnamefont
  {Tuovinen}},\ }\href {\doibase 10.1103/PhysRevB.96.195429} {\bibfield
  {journal} {\bibinfo  {journal} {Phys. Rev. B}\ }\textbf {\bibinfo {volume}
  {96}},\ \bibinfo {pages} {195429} (\bibinfo {year} {2017})}\BibitemShut
  {NoStop}%
\bibitem [{\citenamefont {Talarico}\ \emph {et~al.}(2020)\citenamefont
  {Talarico}, \citenamefont {Maniscalco},\ and\ \citenamefont
  {Lo~Gullo}}]{Talarico2020}%
  \BibitemOpen
  \bibfield  {author} {\bibinfo {author} {\bibfnamefont {N.~W.}\ \bibnamefont
  {Talarico}}, \bibinfo {author} {\bibfnamefont {S.}~\bibnamefont
  {Maniscalco}}, \ and\ \bibinfo {author} {\bibfnamefont {N.}~\bibnamefont
  {Lo~Gullo}},\ }\href {\doibase 10.1103/PhysRevB.101.045103} {\bibfield
  {journal} {\bibinfo  {journal} {Phys. Rev. B}\ }\textbf {\bibinfo {volume}
  {101}},\ \bibinfo {pages} {045103} (\bibinfo {year} {2020})}\BibitemShut
  {NoStop}%
\bibitem [{\citenamefont {Khodas}\ \emph {et~al.}(2007)\citenamefont {Khodas},
  \citenamefont {Pustilnik}, \citenamefont {Kamenev},\ and\ \citenamefont
  {Glazman}}]{Khodas2007}%
  \BibitemOpen
  \bibfield  {author} {\bibinfo {author} {\bibfnamefont {M.}~\bibnamefont
  {Khodas}}, \bibinfo {author} {\bibfnamefont {M.}~\bibnamefont {Pustilnik}},
  \bibinfo {author} {\bibfnamefont {A.}~\bibnamefont {Kamenev}}, \ and\
  \bibinfo {author} {\bibfnamefont {L.~I.}\ \bibnamefont {Glazman}},\ }\href
  {\doibase 10.1103/PhysRevB.76.155402} {\bibfield  {journal} {\bibinfo
  {journal} {Phys. Rev. B - Condens. Matter Mater. Phys.}\ }\textbf {\bibinfo
  {volume} {76}},\ \bibinfo {pages} {1} (\bibinfo {year} {2007})}\BibitemShut
  {NoStop}%
\bibitem [{\citenamefont {Imambekov}\ and\ \citenamefont
  {Glazman}(2008)}]{Imambekov2008}%
  \BibitemOpen
  \bibfield  {author} {\bibinfo {author} {\bibfnamefont {A.}~\bibnamefont
  {Imambekov}}\ and\ \bibinfo {author} {\bibfnamefont {L.~I.}\ \bibnamefont
  {Glazman}},\ }\href {\doibase 10.1103/PhysRevLett.100.206805} {\bibfield
  {journal} {\bibinfo  {journal} {Phys. Rev. Lett.}\ }\textbf {\bibinfo
  {volume} {100}},\ \bibinfo {pages} {1} (\bibinfo {year} {2008})}\BibitemShut
  {NoStop}%
\bibitem [{\citenamefont {Pereira}\ \emph {et~al.}(2008)\citenamefont
  {Pereira}, \citenamefont {White},\ and\ \citenamefont
  {Affleck}}]{Pereira2008}%
  \BibitemOpen
  \bibfield  {author} {\bibinfo {author} {\bibfnamefont {R.~G.}\ \bibnamefont
  {Pereira}}, \bibinfo {author} {\bibfnamefont {S.~R.}\ \bibnamefont {White}},
  \ and\ \bibinfo {author} {\bibfnamefont {I.}~\bibnamefont {Affleck}},\ }\href
  {\doibase 10.1103/PhysRevLett.100.027206} {\bibfield  {journal} {\bibinfo
  {journal} {Phys. Rev. Lett.}\ }\textbf {\bibinfo {volume} {100}},\ \bibinfo
  {pages} {027206} (\bibinfo {year} {2008})}\BibitemShut {NoStop}%
\bibitem [{\citenamefont {Imambekov}\ and\ \citenamefont
  {Glazman}(2009{\natexlab{a}})}]{Imambekov2009}%
  \BibitemOpen
  \bibfield  {author} {\bibinfo {author} {\bibfnamefont {A.}~\bibnamefont
  {Imambekov}}\ and\ \bibinfo {author} {\bibfnamefont {L.~I.}\ \bibnamefont
  {Glazman}},\ }\href {\doibase 10.1126/science.1165403} {\bibfield  {journal}
  {\bibinfo  {journal} {Science (80-. ).}\ }\textbf {\bibinfo {volume} {323}},\
  \bibinfo {pages} {228} (\bibinfo {year} {2009}{\natexlab{a}})}\BibitemShut
  {NoStop}%
\bibitem [{\citenamefont {Imambekov}\ and\ \citenamefont
  {Glazman}(2009{\natexlab{b}})}]{Imambekov2009a}%
  \BibitemOpen
  \bibfield  {author} {\bibinfo {author} {\bibfnamefont {A.}~\bibnamefont
  {Imambekov}}\ and\ \bibinfo {author} {\bibfnamefont {L.~I.}\ \bibnamefont
  {Glazman}},\ }\href {\doibase 10.1103/PhysRevLett.102.126405} {\bibfield
  {journal} {\bibinfo  {journal} {Phys. Rev. Lett.}\ }\textbf {\bibinfo
  {volume} {102}},\ \bibinfo {pages} {12} (\bibinfo {year}
  {2009}{\natexlab{b}})}\BibitemShut {NoStop}%
\bibitem [{\citenamefont {Kamenev}\ and\ \citenamefont
  {Glazman}(2009)}]{Kamenev2009}%
  \BibitemOpen
  \bibfield  {author} {\bibinfo {author} {\bibfnamefont {A.}~\bibnamefont
  {Kamenev}}\ and\ \bibinfo {author} {\bibfnamefont {L.~I.}\ \bibnamefont
  {Glazman}},\ }\href {\doibase 10.1103/PhysRevA.80.011603} {\bibfield
  {journal} {\bibinfo  {journal} {Phys. Rev. A - At. Mol. Opt. Phys.}\ }\textbf
  {\bibinfo {volume} {80}},\ \bibinfo {pages} {011603} (\bibinfo {year}
  {2009})}\BibitemShut {NoStop}%
\bibitem [{\citenamefont {Imambekov}\ \emph {et~al.}(2012)\citenamefont
  {Imambekov}, \citenamefont {Schmidt},\ and\ \citenamefont
  {Glazman}}]{Imambekov2012}%
  \BibitemOpen
  \bibfield  {author} {\bibinfo {author} {\bibfnamefont {A.}~\bibnamefont
  {Imambekov}}, \bibinfo {author} {\bibfnamefont {T.~L.}\ \bibnamefont
  {Schmidt}}, \ and\ \bibinfo {author} {\bibfnamefont {L.~I.}\ \bibnamefont
  {Glazman}},\ }\href {\doibase 10.1103/RevModPhys.84.1253} {\bibfield
  {journal} {\bibinfo  {journal} {Rev. Mod. Phys.}\ }\textbf {\bibinfo {volume}
  {84}},\ \bibinfo {pages} {1253} (\bibinfo {year} {2012})}\BibitemShut
  {NoStop}%
\bibitem [{\citenamefont {Ristivojevic}(2014)}]{Ristivojevic2014}%
  \BibitemOpen
  \bibfield  {author} {\bibinfo {author} {\bibfnamefont {Z.}~\bibnamefont
  {Ristivojevic}},\ }\href {\doibase 10.1103/PhysRevLett.113.015301} {\bibfield
   {journal} {\bibinfo  {journal} {Phys. Rev. Lett.}\ }\textbf {\bibinfo
  {volume} {113}},\ \bibinfo {pages} {0} (\bibinfo {year} {2014})}\BibitemShut
  {NoStop}%
\bibitem [{\citenamefont {Markhof}\ and\ \citenamefont
  {Meden}(2016)}]{Markhof2016}%
  \BibitemOpen
  \bibfield  {author} {\bibinfo {author} {\bibfnamefont {L.}~\bibnamefont
  {Markhof}}\ and\ \bibinfo {author} {\bibfnamefont {V.}~\bibnamefont
  {Meden}},\ }\href {\doibase 10.1103/PhysRevB.93.085108} {\bibfield  {journal}
  {\bibinfo  {journal} {Phys. Rev. B}\ }\textbf {\bibinfo {volume} {93}},\
  \bibinfo {pages} {1} (\bibinfo {year} {2016})}\BibitemShut {NoStop}%
\bibitem [{\citenamefont {Campbell}\ and\ \citenamefont
  {Gangardt}(2017)}]{Campbell2017}%
  \BibitemOpen
  \bibfield  {author} {\bibinfo {author} {\bibfnamefont {A.}~\bibnamefont
  {Campbell}}\ and\ \bibinfo {author} {\bibfnamefont {D.}~\bibnamefont
  {Gangardt}},\ }\href {\doibase 10.21468/SciPostPhys.3.2.015} {\bibfield
  {journal} {\bibinfo  {journal} {SciPost Phys.}\ }\textbf {\bibinfo {volume}
  {3}},\ \bibinfo {pages} {015} (\bibinfo {year} {2017})}\BibitemShut {NoStop}%
\bibitem [{\citenamefont {Kozlowski}\ \emph {et~al.}(2011)\citenamefont
  {Kozlowski}, \citenamefont {Maillet},\ and\ \citenamefont
  {Slavnov}}]{Kozlowski2011}%
  \BibitemOpen
  \bibfield  {author} {\bibinfo {author} {\bibfnamefont {K.~K.}\ \bibnamefont
  {Kozlowski}}, \bibinfo {author} {\bibfnamefont {J.~M.}\ \bibnamefont
  {Maillet}}, \ and\ \bibinfo {author} {\bibfnamefont {N.~A.}\ \bibnamefont
  {Slavnov}},\ }\href {\doibase 10.1088/1742-5468/2011/03/P03019} {\bibfield
  {journal} {\bibinfo  {journal} {J. Stat. Mech. Theory Exp.}\ }\textbf
  {\bibinfo {volume} {2011}},\ \bibinfo {pages} {P03019} (\bibinfo {year}
  {2011})}\BibitemShut {NoStop}%
\bibitem [{\citenamefont {Talarico}\ \emph {et~al.}(2019)\citenamefont
  {Talarico}, \citenamefont {Maniscalco},\ and\ \citenamefont
  {Lo~Gullo}}]{Talarico2019}%
  \BibitemOpen
  \bibfield  {author} {\bibinfo {author} {\bibfnamefont {N.~W.}\ \bibnamefont
  {Talarico}}, \bibinfo {author} {\bibfnamefont {S.}~\bibnamefont
  {Maniscalco}}, \ and\ \bibinfo {author} {\bibfnamefont {N.}~\bibnamefont
  {Lo~Gullo}},\ }\href {\doibase 10.1002/pssb.201800501} {\bibfield  {journal}
  {\bibinfo  {journal} {physica status solidi (b)}\ }\textbf {\bibinfo {volume}
  {256}},\ \bibinfo {pages} {1800501} (\bibinfo {year} {2019})}\BibitemShut
  {NoStop}%
\bibitem [{\citenamefont {Lo~Gullo}\ and\ \citenamefont
  {Dell'Anna}(2015)}]{LoGullo2015}%
  \BibitemOpen
  \bibfield  {author} {\bibinfo {author} {\bibfnamefont {N.}~\bibnamefont
  {Lo~Gullo}}\ and\ \bibinfo {author} {\bibfnamefont {L.}~\bibnamefont
  {Dell'Anna}},\ }\href {\doibase 10.1103/PhysRevA.92.063619} {\bibfield
  {journal} {\bibinfo  {journal} {Phys. Rev. A}\ }\textbf {\bibinfo {volume}
  {92}},\ \bibinfo {pages} {063619} (\bibinfo {year} {2015})}\BibitemShut
  {NoStop}%
\bibitem [{\citenamefont {Lo~Gullo}\ and\ \citenamefont
  {Dell'Anna}(2016)}]{LoGullo2016}%
  \BibitemOpen
  \bibfield  {author} {\bibinfo {author} {\bibfnamefont {N.}~\bibnamefont
  {Lo~Gullo}}\ and\ \bibinfo {author} {\bibfnamefont {L.}~\bibnamefont
  {Dell'Anna}},\ }\href {\doibase 10.1103/PhysRevB.94.184308} {\bibfield
  {journal} {\bibinfo  {journal} {Phys. Rev. B}\ }\textbf {\bibinfo {volume}
  {94}},\ \bibinfo {pages} {184308} (\bibinfo {year} {2016})}\BibitemShut
  {NoStop}%
\bibitem [{\citenamefont {Settino}\ \emph {et~al.}(2020)\citenamefont
  {Settino}, \citenamefont {Talarico}, \citenamefont {Cosco}, \citenamefont
  {Plastina}, \citenamefont {Maniscalco},\ and\ \citenamefont
  {Lo~Gullo}}]{Settino2020}%
  \BibitemOpen
  \bibfield  {author} {\bibinfo {author} {\bibfnamefont {J.}~\bibnamefont
  {Settino}}, \bibinfo {author} {\bibfnamefont {N.~W.}\ \bibnamefont
  {Talarico}}, \bibinfo {author} {\bibfnamefont {F.}~\bibnamefont {Cosco}},
  \bibinfo {author} {\bibfnamefont {F.}~\bibnamefont {Plastina}}, \bibinfo
  {author} {\bibfnamefont {S.}~\bibnamefont {Maniscalco}}, \ and\ \bibinfo
  {author} {\bibfnamefont {N.}~\bibnamefont {Lo~Gullo}},\ }\href {\doibase
  10.1103/PhysRevB.101.144303} {\bibfield  {journal} {\bibinfo  {journal}
  {Phys. Rev. B}\ }\textbf {\bibinfo {volume} {101}},\ \bibinfo {pages}
  {144303} (\bibinfo {year} {2020})}\BibitemShut {NoStop}%
\bibitem [{\citenamefont {Lieb}\ and\ \citenamefont
  {Liniger}(1963)}]{Lieb1963}%
  \BibitemOpen
  \bibfield  {author} {\bibinfo {author} {\bibfnamefont {E.~H.}\ \bibnamefont
  {Lieb}}\ and\ \bibinfo {author} {\bibfnamefont {W.}~\bibnamefont {Liniger}},\
  }\href {\doibase 10.1103/PhysRev.130.1605} {\bibfield  {journal} {\bibinfo
  {journal} {Phys. Rev.}\ }\textbf {\bibinfo {volume} {130}},\ \bibinfo {pages}
  {1605} (\bibinfo {year} {1963})}\BibitemShut {NoStop}%
\bibitem [{\citenamefont {Lieb}(1963)}]{Lieb1963a}%
  \BibitemOpen
  \bibfield  {author} {\bibinfo {author} {\bibfnamefont {E.~H.}\ \bibnamefont
  {Lieb}},\ }\href {\doibase 10.1103/PhysRev.130.1616} {\bibfield  {journal}
  {\bibinfo  {journal} {Phys. Rev.}\ }\textbf {\bibinfo {volume} {130}},\
  \bibinfo {pages} {1616} (\bibinfo {year} {1963})}\BibitemShut {NoStop}%
\bibitem [{\citenamefont {Yang}\ and\ \citenamefont {Yang}(1969)}]{Yang1969}%
  \BibitemOpen
  \bibfield  {author} {\bibinfo {author} {\bibfnamefont {C.~N.}\ \bibnamefont
  {Yang}}\ and\ \bibinfo {author} {\bibfnamefont {C.~P.}\ \bibnamefont
  {Yang}},\ }\href {\doibase 10.1063/1.1664947} {\bibfield  {journal} {\bibinfo
   {journal} {J. Math. Phys.}\ }\textbf {\bibinfo {volume} {10}},\ \bibinfo
  {pages} {1115} (\bibinfo {year} {1969})}\BibitemShut {NoStop}%
\bibitem [{\citenamefont {Cazalilla}\ \emph {et~al.}(2011)\citenamefont
  {Cazalilla}, \citenamefont {Citro}, \citenamefont {Giamarchi}, \citenamefont
  {Orignac},\ and\ \citenamefont {Rigol}}]{Cazalilla2011}%
  \BibitemOpen
  \bibfield  {author} {\bibinfo {author} {\bibfnamefont {M.~A.}\ \bibnamefont
  {Cazalilla}}, \bibinfo {author} {\bibfnamefont {R.}~\bibnamefont {Citro}},
  \bibinfo {author} {\bibfnamefont {T.}~\bibnamefont {Giamarchi}}, \bibinfo
  {author} {\bibfnamefont {E.}~\bibnamefont {Orignac}}, \ and\ \bibinfo
  {author} {\bibfnamefont {M.}~\bibnamefont {Rigol}},\ }\href {\doibase
  10.1103/RevModPhys.83.1405} {\bibfield  {journal} {\bibinfo  {journal} {Rev.
  Mod. Phys.}\ }\textbf {\bibinfo {volume} {83}},\ \bibinfo {pages} {1405}
  (\bibinfo {year} {2011})}\BibitemShut {NoStop}%
\bibitem [{\citenamefont {Mahan}(1967)}]{Mahan1967}%
  \BibitemOpen
  \bibfield  {author} {\bibinfo {author} {\bibfnamefont {G.~D.}\ \bibnamefont
  {Mahan}},\ }\href {\doibase 10.1103/PhysRev.163.612} {\bibfield  {journal}
  {\bibinfo  {journal} {Phys. Rev.}\ }\textbf {\bibinfo {volume} {163}},\
  \bibinfo {pages} {612} (\bibinfo {year} {1967})}\BibitemShut {NoStop}%
\bibitem [{\citenamefont {Nozi{\`{e}}res}\ and\ \citenamefont {{De
  Dominicis}}(1969)}]{Nozieres1969a}%
  \BibitemOpen
  \bibfield  {author} {\bibinfo {author} {\bibfnamefont {P.}~\bibnamefont
  {Nozi{\`{e}}res}}\ and\ \bibinfo {author} {\bibfnamefont {C.~T.}\
  \bibnamefont {{De Dominicis}}},\ }\href {\doibase 10.1103/PhysRev.178.1097}
  {\bibfield  {journal} {\bibinfo  {journal} {Phys. Rev.}\ }\textbf {\bibinfo
  {volume} {178}},\ \bibinfo {pages} {1097} (\bibinfo {year}
  {1969})}\BibitemShut {NoStop}%
\bibitem [{\citenamefont {Sindona}\ \emph {et~al.}(2013)\citenamefont
  {Sindona}, \citenamefont {Goold}, \citenamefont {{Lo Gullo}}, \citenamefont
  {Lorenzo},\ and\ \citenamefont {Plastina}}]{Sindona2013}%
  \BibitemOpen
  \bibfield  {author} {\bibinfo {author} {\bibfnamefont {A.}~\bibnamefont
  {Sindona}}, \bibinfo {author} {\bibfnamefont {J.}~\bibnamefont {Goold}},
  \bibinfo {author} {\bibfnamefont {N.}~\bibnamefont {{Lo Gullo}}}, \bibinfo
  {author} {\bibfnamefont {S.}~\bibnamefont {Lorenzo}}, \ and\ \bibinfo
  {author} {\bibfnamefont {F.}~\bibnamefont {Plastina}},\ }\href {\doibase
  10.1103/PhysRevLett.111.165303} {\bibfield  {journal} {\bibinfo  {journal}
  {Phys. Rev. Lett.}\ }\textbf {\bibinfo {volume} {111}},\ \bibinfo {pages}
  {165303} (\bibinfo {year} {2013})}\BibitemShut {NoStop}%
\bibitem [{\citenamefont {Sindona}\ \emph {et~al.}(2015)\citenamefont
  {Sindona}, \citenamefont {Pisarra}, \citenamefont {Gravina}, \citenamefont
  {{Vacacela Gomez}}, \citenamefont {Riccardi}, \citenamefont {Falcone},\ and\
  \citenamefont {Plastina}}]{Sindona2015}%
  \BibitemOpen
  \bibfield  {author} {\bibinfo {author} {\bibfnamefont {A.}~\bibnamefont
  {Sindona}}, \bibinfo {author} {\bibfnamefont {M.}~\bibnamefont {Pisarra}},
  \bibinfo {author} {\bibfnamefont {M.}~\bibnamefont {Gravina}}, \bibinfo
  {author} {\bibfnamefont {C.}~\bibnamefont {{Vacacela Gomez}}}, \bibinfo
  {author} {\bibfnamefont {P.}~\bibnamefont {Riccardi}}, \bibinfo {author}
  {\bibfnamefont {G.}~\bibnamefont {Falcone}}, \ and\ \bibinfo {author}
  {\bibfnamefont {F.}~\bibnamefont {Plastina}},\ }\href {\doibase
  10.3762/bjnano.6.78} {\bibfield  {journal} {\bibinfo  {journal} {Beilstein J.
  Nanotechnol.}\ }\textbf {\bibinfo {volume} {6}},\ \bibinfo {pages} {755}
  (\bibinfo {year} {2015})}\BibitemShut {NoStop}%
\bibitem [{\citenamefont {Olshanii}\ and\ \citenamefont
  {Dunjko}(2003)}]{Olshanii2004}%
  \BibitemOpen
  \bibfield  {author} {\bibinfo {author} {\bibfnamefont {M.}~\bibnamefont
  {Olshanii}}\ and\ \bibinfo {author} {\bibfnamefont {V.}~\bibnamefont
  {Dunjko}},\ }\href {\doibase 10.1103/PhysRevLett.91.090401} {\bibfield
  {journal} {\bibinfo  {journal} {Phys. Rev. Lett.}\ }\textbf {\bibinfo
  {volume} {91}},\ \bibinfo {pages} {090401} (\bibinfo {year}
  {2003})}\BibitemShut {NoStop}%
\end{thebibliography}%

\newpage
\onecolumngrid
\appendix

\section*{Calculation of lesser and greater Green's functions}
\label{app:lessgreat}
\subsection{Lesser Green's function: $G^< (x,t,y,t')$}
\label{sec:grLes}
We provide here the details of  the calculation for the lesser Green's function $G^<(x,t,y,t')$ for a $N$-particle TG gas.
The lesser Green's function is defined as
\begin{equation}
\begin{split}\label{eq:grLesI}
\imath G^< (x,t,y,t')_{\boldsymbol \eta}&= \expval{\hat \psi^\dagger (y,t') \hat \psi(x,t)}_{\boldsymbol \eta} \\
&=  \expval{ e^{i H t'} \hat \psi^\dagger (y) e^{-i H t'} e^{i H t} \hat \psi(x) e^{-i H t}}_{\boldsymbol \eta}
\end{split} 
\end{equation} 
where $\langle...\rangle_{\boldsymbol \eta}$ indicates the expectation value over the many-body state $|{\boldsymbol \eta}\rangle$, $H$ is the many-body Hamiltonian and $\hat \psi(x)$, $\hat \psi^\dagger(x) $ are bosonic field operators, satisfying the communtation relations $[\hat \psi(x),\hat \psi^\dagger(y)]=\delta(x-y)$.

In order to perform the exact calculation for a TG gas, based on the Girardeau mapping on noninteracting fermions, it is useful to rewrite  the Green's function in the first quantization formalism. We introduce the completeness relation in the $N-1$ particles Hilbert space $\sum_n \dyad{n} = \mathbb{I}_{N-1}$, with $\ket{n}$ being an eigenstate of the TG Hamiltonian and the sum being restricted to inequivalent states, and the completeness relation in the $N-1$ particles Hilbert space in the position representation $\frac{1}{{N-1!}} \int \dd X \dyad{X}= \mathbb{I}_{N-1}$, with $X=x_2 \dots x_N$.  
\begin{equation}
\begin{split}
\imath G^< (x,t,y,t')_{\boldsymbol \eta} & = \frac{1}{(N-1!)^2}\expval{e^{i H t'} \hat \psi^\dagger (y) \int \dd Y \dyad{Y} e^{-i H t'} (\sum_n \dyad{n}) e^{i H t}   \int \dd X \dyad{X} \hat \psi(x)  e^{-i H t}}_{\boldsymbol \eta} \\
&=\frac{1}{(N-1!)^2} \sum_n \int dY \int dX 
{}_{t\!'}\!\!\braket{\boldsymbol \eta}{y,Y} \braket{Y}{n}_{t\!'} {}_{t}\!\braket{n}{X} \braket{x,X}{\boldsymbol \eta}_{t}\\
&= \frac{1}{(N-1!)^2} \sum_n \int dY \Psi_{\boldsymbol \eta}^*(y,Y;t')\Psi_n(Y;t') \int dX \Psi_n^*(X;t) \Psi_{\boldsymbol \eta}(x,X;t),
\end{split} 
\end{equation} 
where we have used the definition of many-body wavefunction $\braket{x,X}{\boldsymbol \eta}=\Psi_{\boldsymbol \eta}(x,X)$ and similarly  $\braket{X}{n}=\Psi_n(X)$.
We now apply the Bose-Fermi mapping and write the bosonic wavefunction $\Psi_{\boldsymbol \eta}(x_1,...x_N)=\prod_{j,\ell\in\{\vec {\boldsymbol \eta}\}} {\rm sign}(x_j-x_\ell) \Psi_{\boldsymbol \eta}^F(x_1,..,x_N)$, where  $\Psi_{\boldsymbol \eta}^F(x_1,..,x_N)=(1/\sqrt{N!})\det[\phi_{{\boldsymbol \eta}_j}(x_\ell)]$ with $j,\ell=1..N$, $\phi_j(x)$ the single-particle orbitals for the given external potential with energy $e_j$,  and we have introduced the notation  ${\boldsymbol \eta}=\{\eta_1,...\eta_N \}$. We  thus obtain the expression for the lesser Green's function of a TG gas: 
\begin{equation}\label{eq:lesser3}
\imath G^< (x,t,y,t')_{\boldsymbol \eta} = \frac{1}{(N-1!)^2} \sum_n  \int dX \prod_{k=2}^N {\rm sign}(x-x_k) \Psi_{\boldsymbol \eta}^F(x,X;t)\Psi_n^{*F}(X;t) 
\int dY \prod_{k=2}^N {\rm sign}(y-y_k) \Psi_{\boldsymbol \eta}^{*F}(y,Y;t')\Psi_n^{F}(Y;t')
\end{equation} 
Each of the two multidimensional integrals can be evaluated separately; we will start by writing the first one as a function of single particle states, by using the properties of Slater determinants. 
We identify the generic $(N-1)$ particles eigenstate of the free fermions Hamiltonian, labeled by $n$, as the one with single-particle orbitals $\boldsymbol \alpha = \{\alpha_2, ... ,\alpha_N \}$. 
Expanding the determinant in $\Psi_{\boldsymbol \eta}^F(x,X;t)$ by the first column, we have
	\begin{equation}
	\label{eq:split}
	\begin{split}
	&\int dX \prod_{k=2}^N {\rm sign}(x-x_k) \Psi_{\boldsymbol \eta}^F(x,X;t)\Psi_n^{*F}(X;t)= \\
	&= \sum_{i=1}^N (-1)^{i+1}\phi_{\eta_i}(x,t) \int dX \prod_{k=2}^N {\rm sign}(x-x_k)  
	\smdet{\phi_{\eta_1}(x_2,t) & \dots & \phi_{\eta_1}(x_N,t)\\
		\vdots & \dots & \vdots\\
		\phi_{\eta_{i-1}}(x_2,t) & \dots & \phi_{\eta_{i-1}}(x_N,t)\\
		\phi_{\eta_{i+1}}(x_2,t) & \dots & \phi_{\eta_{i+1}}(x_N,t)\\
		\vdots & \vdots & \vdots\\
		\phi_{\eta_N}(x_2,t) & \dots & \phi_{\eta_N}(x_N,t)\\}
	\smdet{\phi^*{\alpha_2}(x_2,t) & \dots & \phi^*{\alpha_2}(x_N,t)\\
		\vdots & \ddots & \vdots\\
		\phi^*{\alpha_N}(x_2,t) & \dots & \phi^*{\alpha_N}(x_N,t)\\}
	\end{split}
	\end{equation}
%
We can combine the two determinants using the Andréief's integration formula \cite{Forrester2002}

\begin{equation}\label{eq:Andreief}
\int \dd x_1 \dots \int \dd x_M \det[f_j(x_k]_{j,k=1,M} \det[g_j(x_k]_{j,k=1,M} = M! \det[\int \dd x f_j(x) g_k(x) ]_{j,k=1,M}                                                                                                                                                                                                                                                                                                                                                                                                                                                                                                    \end{equation}

Then, noticing the fact that $\int_{-\infty}^{\infty} {\rm sign}(x-\bar x ) f(\bar x) \dd \bar x = \int_{-\infty}^{\infty}f(\bar x ) \dd \bar x - 2 \int_{x}^{\infty} f (\bar x) \dd \bar x $, we obtain
	\begin{equation}
	\int \dd X \prod_{k=2}^N {\rm sign}(x-x_k) \Psi_{\boldsymbol \eta}^F(x,X;t)\Psi_n^{*F}(X;t)= (N-1)!\sum_{i=1}^N (-1)^{i+1} \phi_{\eta_i}(x,t) 
	\det[\textbf P (x,t)]_{{\boldsymbol \eta} \smallsetminus  \{\eta_i\},{\boldsymbol \alpha}}.
	\end{equation}
	The determinant $\det [ \textbf P]_{{\boldsymbol \eta} \smallsetminus  \{\eta_i\},{\boldsymbol \alpha}}$ is the $N-1$ order minor of the matrix $ \textbf P$ having selected the rows ${\boldsymbol \eta}\smallsetminus  \{\eta_i\}$ and the columns ${\boldsymbol \alpha}$, and
	\begin{multline}\label{eq:TGP}
	P_{l,m}(x,t)= \int_{-\infty}^{\infty} \phi_l(\bar x,t)\phi^*_m(\bar x,t) \dd \bar x - 2 \int_{x}^{\infty} \phi_l(\bar x,t)\phi^*_m(\bar x,t) \dd \bar x=\delta_{l,m} - 2 \ e^{-\imath t (e_l - e_m)} \int_{x}^{\infty} \phi_l(\bar x)\phi^*_m(\bar x) \dd \bar x
	\end{multline}
In the same way we can write the second integral in the expression for $G^<$, obtaining:
\begin{equation}
\imath G^< (x,t,y,t')_\eta=\widetilde \sum_{\boldsymbol \alpha}  \sum_{i,j=1}^N (-1)^{i+j} \phi_{\eta_i} (x,t) \phi^*_{\eta_j} (y,t')
\det[\textbf P(x,t)]_{{\boldsymbol \eta}\smallsetminus  \{\eta_i\},{\boldsymbol \alpha}} \det[\textbf P(y,t')]_{{\boldsymbol \alpha},{\boldsymbol \eta}\smallsetminus  \{\eta_j\}}
\end{equation}

The sum over $n$ of Eq.~(\ref{eq:lesser3}) corresponds to the sum over $\boldsymbol \alpha$ in the equation above, that has to be restricted to collections of indices that are not related by permutations, and that will be indicated from now on by $ \widetilde \sum$. This sum can be simplified by using the generalized Cauchy-Binet formula for the product of minors
\begin{equation}
\widetilde \sum_{\boldsymbol \alpha}\ \det[\textbf A]_{\vec I,{\boldsymbol \alpha}} \det[\textbf B]_{{\boldsymbol \alpha},\vec J} = \det[\textbf {A B}]_{\vec I,\vec J}                                                                                                                                                                                                                                                                                                                                                                                                                                                                                                                                                                   \end{equation}
obtaining:
\begin{multline}
(-1)^{i+j} \sum_{\boldsymbol \alpha} \det[\textbf P(x,t)]_{{\boldsymbol \eta}\smallsetminus  \{\eta_i\},{\boldsymbol \alpha}} \det[\textbf P(y,t')]_{{\boldsymbol \alpha},{\boldsymbol \eta}\smallsetminus  \{\eta_j\}}= \\
(-1)^{i+j} \det[\textbf P(x,t) \textbf P(y,t')]_{{{\boldsymbol \eta}\smallsetminus  \{\eta_i\}},{{\boldsymbol \eta}\smallsetminus  \{\eta_j\}}}
={\{[{\textbf  P} (x,t) {\textbf  P} (y,t') ]_{{\boldsymbol \eta},{\boldsymbol \eta}}\}^{-1}}^T \det[{\textbf  P} (x,t) {\textbf  P} (y,t')]_{{\boldsymbol \eta},{\boldsymbol \eta}}
\end{multline}
where, in the last step, we have used the definition of the inverse of a matrix via minors. It is important to note that the product between matrices in the last equation is not constrained to the ${\boldsymbol \eta}$ elements of the single particle Hilbert space; rather, it spans the whole single particle Hilbert space. In the numerical calculation a suitably chosen truncation has been employed.

We can finally write:
\begin{equation}
\imath G^< (x,t,y,t')_{\boldsymbol \eta}= \sum_{i,j=1}^N \phi_{\eta_i} (x) e^{- \imath e_{\eta_i} t} \phi^*_{\eta_j} 
(y) e^{ \imath e_{\eta_j} t'} A_{\eta_i,\eta_j} (x,t,y,t')
\end{equation} 
with
\begin{equation}
{\textbf  A}_{{\boldsymbol \eta},{\boldsymbol \eta}} (x,t,y,t')={\{[{\textbf  P} (x,t) {\textbf  P} (y,t') ]_{{\boldsymbol \eta},{\boldsymbol \eta}}\}^{-1}}^T \det[{\textbf  P} (x,t) {\textbf  P} (y,t')]_{{\boldsymbol \eta},{\boldsymbol \eta}}.
\end{equation}
This result generalizes the calculation of the one-body density matrix in  Ref.~\cite{Pezer2007}, to which it reduces when $\ket {\boldsymbol \eta}$ corresponds to the ground state and when we take equal times $t=t'$.

\subsection{Greater Green's Function: $G^>(x,t,y,t')$}
\label{sec:grGre}
In an analogous fashion, we can evaluate the greater Green's function $G^>(x,t,y,t')$ for a TG gas, which is defined as
\begin{equation}
\label{eq:grGreI}
\imath G^> (x,t,y,t')_{\boldsymbol \eta}= \expval{\hat \psi(x,t) \hat \psi^\dagger (y,t')}_{\boldsymbol \eta} 
=  \expval{ e^{i H t} \hat \psi(x) e^{-i H t}e^{i H t'} \hat \psi^\dagger (y) e^{-i H t'}}_{\boldsymbol \eta}
\end{equation} 
In order to write it in the first quantization formalism and to apply the time evolution operator, we introduce this time the completeness relation in the $N+1$ particles Hilbert space $\sum_n \dyad{n} = \mathbb{I}_{N+1}$, with $\ket{n}$ being an eigenstate of the TG Hamiltonian with $N+1$ particles.
The expression for the greater Green's function for a TG gas then reads
\bigbreak
\begin{equation}
\begin{split}\label{eq:grGreII}
\hspace*{-20pt} \imath G^> (x,t,y,t')_{\boldsymbol \eta} 
&= \frac{1}{(N!)^2}\expval{ e^{i H t}  \int dX \dyad{X} \hat \psi(x) e^{-i H t} (\sum_n \dyad{n}) e^{i H t'} \hat \psi^\dagger (y) \int dY \dyad{Y} e^{-i H t'} }_{\boldsymbol \eta}\\
&= \frac{1}{(N!)^2}\sum_n \int dX \int dY 
{}_{t}\!\!\braket{{\boldsymbol \eta}}{X} \braket{x,X}{n}_{\!t} {}_{t\!'}\!\!\braket{n}{y,Y} \braket{Y}{\boldsymbol \eta}_{\!t\!'}=\\
&= \frac{1}{(N!)^2}\sum_n \int dX \Psi_{\boldsymbol \eta}^*(X;t)\Psi_n(x,X;t) \int dY \Psi_n^*(y,Y;t') \Psi_{\boldsymbol \eta}(Y;t')
\end{split} 
\end{equation} 
The use of the Bose-Fermi mapping then leads to 
\begin{equation}
\begin{split}
\imath G^> (x,t,y,t')_{\boldsymbol \eta} =\frac{1}{(N!)^2} \sum_n \int \dd X \prod_{k=1}^N {\rm sign}(x-x_k) \Psi^{F*}_{\boldsymbol \eta}(X;t)\Psi^{F}_n(x,X;t) \\
\times \int \dd Y \prod_{k=1}^N {\rm sign}(y-y_k) \Psi^{F}_{\boldsymbol \eta}(Y;t')\Psi^{F*}_n(y,Y;t').
\end{split}
\end{equation}
As in  Eq.~\ref{eq:split}, the calculation of the first integral yields
\begin{equation}
\begin{split}
&\int \dd X \prod_{k=1}^N {\rm sign}(x-x_k) \Psi_{\boldsymbol \eta}^*(X;t)\Psi_n(x,X;t)\\
=& \sum_{i=1}^{N+1} {(-1)^{i+1}}\phi_{\alpha_i}(x,t) \int \dd X \prod_{k=1}^N {\rm sign}(x-x_k) \\
&\hspace*{100pt} \times \smdet{\phi^*_{\eta_1}(x_1,t) & \dots & \phi^*_{\eta_1}(x_N,t)\\
	\vdots & \ddots & \vdots\\
	\phi^*_{\eta_N}(x_1,t) & \dots & \phi^*_{\eta_N}(x_N,t)\\}
\smdet{\phi_{\alpha_1}(x_1,t) & \dots & \phi_{\alpha_i}(x_{N+1},t)\\
	\vdots & \dots & \vdots\\
	\phi_{\alpha_{i-1}}(x_1,t) & \dots & \phi_{\alpha_{i-1}}(x_{N+1},t)\\
	\phi_{\alpha_{i+1}}(x_1,t) & \dots & \phi_{\alpha_{i+1}}(x_{N+1},t)\\
	\vdots & \vdots & \vdots\\
	\phi_{\alpha_{N+1}}(x_1,t) & \dots & \phi_{\alpha_{N+1}}(x_{N+1},t)\\}.
\end{split}
\end{equation}
As for the lesser Green's function, we can combine the two determinants using the Andréief's integration formula, Eq.~(\ref{eq:Andreief}). Then, by noticing that $\int_{-\infty}^{\infty} {\rm sign}(x-\bar x ) f(\bar x) \dd \bar x = \int_{-\infty}^{\infty}f(\bar x ) \dd \bar x - 2 \int_{x}^{\infty} f (\bar x) \dd \bar x $, we obtain
\begin{equation}
\int \dd X \prod_{k=1}^N {\rm sign}(x-x_k) \Psi^{F*}_{\boldsymbol \eta}(X;t)\Psi^F_n(x,X;t)= N! \sum_{i=1}^{N+1} (-1)^{i+1} \phi_{\alpha_i}(x,t) 
\ det[\textbf P (x,t)]_{{\boldsymbol \alpha \smallsetminus  \{\alpha_i\}},{\boldsymbol \eta} }.
\end{equation}
The determinant $\det[\textbf P (x,t)]_{{\boldsymbol \alpha \smallsetminus  \{\alpha_i\}},{{\boldsymbol \eta}} }$ is the $N$ order minor of the matrix $ \textbf P$, once we select the rows ${{ \boldsymbol \alpha \smallsetminus  \{\alpha_i\}}}$  and the columns ${\boldsymbol \eta}$,  and $P_{l,m}(x,t)$ defined as in Eq.~(\ref{eq:TGP}).
The main difference with the calculation of the lesser Green's function is that  the Cauchy-Binet theorem cannot be applied to the above expression. We then  insert all the $ \phi_{\alpha_i}(x,t)$ elements into an extended "P" matrix, by adding a "0" column, as  follows,
\begin{equation}
\int dX \prod_{k=1}^N {\rm sign}(x-x_k) \Psi^{F*}_{\boldsymbol \eta}(X;t)\Psi^{F*}_n(x,X;t)=N! \det[\vec \phi(x,t),\textbf P (x,t)]_{{\boldsymbol \alpha},{\{0\} \cup {\boldsymbol \eta}} }, 
\end{equation}
in which we have defined a column vector $\vec \phi(x,t)= [\phi_1(x,t),\dots,\phi_M(x,t)]^T$ on the whole Hilbert space.
Following the same line for the second integral, we obtain:
\begin{equation}
\imath G^> (x,t,y,t')_{\boldsymbol \eta}=\widetilde\sum_{\boldsymbol \alpha} 
\det\mqty[\vec \phi(y,t')^\dagger \\
\textbf P (y,t')]_{{\{0\} \cup {\boldsymbol \eta}},{\boldsymbol \alpha}} 
\det\mqty[\vec \phi(x,t) && \textbf P (x,t)]_{{\boldsymbol \alpha},{\{0\} \cup {\boldsymbol \eta}} }  
\end{equation} 
The sum $\widetilde\sum_{\boldsymbol \alpha}$ has to be restricted to collections of indices that are not related by permutations.
Now we can apply the generalized Cauchy-Binet formula, for products between determinats, obtaining:
\begin{equation}
\begin{split}
\imath G^> (x,t,y,t')_{\boldsymbol \eta}&= 
\det \mqty[\vec \phi(y,t')^\dagger \vec \phi(x,t) && \vec \phi(y,t')^\dagger \textbf P (x,t)\\
\textbf P (y,t') \vec\phi(x,t) && \textbf P (y,t')\textbf P (x,t)]_{{\{0\} \cup {\boldsymbol \eta}},{\{0\} \cup {\boldsymbol \eta}}}\\
&=  \det[\textbf P (y,t')\textbf P (x,t)]_{{\boldsymbol \eta},{\boldsymbol \eta}} \\
&\times \left( \vec \phi(y,t')^\dagger \vec\phi(x,t) -[\vec \phi(y,t')^\dagger \textbf P (x,t)]_{1,{\boldsymbol \eta}}\ [\textbf P (y,t')\textbf P (x,t)]^{-1}_{{\boldsymbol \eta},{\boldsymbol \eta}} \ [\textbf P (y,t') \vec\phi(x,t)]_{{\boldsymbol \eta},1}   \right)
\end{split}
\end{equation}
in which all the products, where not explicitely indicated, should be thought as in the whole Hilbert space.

\subsection{Final expressions for the lesser and greater Green's functions of a TG gas}
\label{app:finallg}

The lesser and greater Green's functions for an eingenstate ${\boldsymbol \eta}$ of the TG Hamiltonian can be finally recast as:
\begin{subequations}
	\begin{equation}\label{eq:lesserTGSupp}
	\imath G^< (x,t,y,t')_{\boldsymbol \eta}=  \det[\textbf P (x,t)\textbf P (y,t')|_{{\boldsymbol \eta}{\boldsymbol \eta}}] a^<(x,t,y,t')
	\end{equation}
	\begin{equation}\label{eq:greaterTGSupp}
	\imath G^> (x,t,y,t')_{\boldsymbol \eta}= \det[\textbf P (y,t')\textbf P (x,t)|_{{\boldsymbol \eta}{\boldsymbol \eta}}] a^>(x,t,y,t')
	\end{equation}
\end{subequations}
with
\begin{subequations} 
	\begin{align} 
	a^<(x,t,y,t') &= {\vec \phi(x,t)_{{\boldsymbol \eta}}^T} \  {[{\textbf P}(x,t) {\textbf P}(y,t')]^{-1 T}}|_{{\boldsymbol \eta}{\boldsymbol \eta}} \   {\vec \phi^*(y,t')}_{{\boldsymbol \eta}}\\
	\begin{split}
	a^>(x,t,y,t') &= \vec \phi(y,t')^\dagger \vec \phi(x,t) -[\vec \phi(y,t')^\dagger \textbf P (x,t)]_{\boldsymbol \eta}\ \\& [\textbf P (y,t')\textbf P (x,t)]^{-1}|_{{\boldsymbol \eta}{\boldsymbol \eta}} \ [\textbf P (y,t') \vec\phi(x,t)]_{{\boldsymbol \eta}}   .
	\end{split}
	\end{align}
\end{subequations} 


From the above expressions we readily  recover the limit of non-interacting fermions by replacing $\text{{\rm sign}}(x-y)$ with  $1$, obtaining $P_{l,m}(x,t)=\delta_{l,m}$, hence
$G^<_{F} (x,t,y,t')_{\boldsymbol \eta} = \imath \sum_{\boldsymbol \eta} e^{ \imath e_i t'}\phi^*_i (y) \phi_i (x) e^{- \imath e_i t} $ and
$G^>_{F} (x,t,y,t')_{\boldsymbol \eta} = -\imath \sum_{\bar{\boldsymbol \eta}}  e^{ \imath e_i t'}\phi^*_i (y) \phi_i (x) e^{- \imath e_i t}$, corresponding to 
the Green's functions  for a non-interacting Fermi gas in the state  ${\boldsymbol \eta}$ \cite{Stefanucci2010}.

\section*{Power-law exponents of the spectral function of a homogeneous Bose  gas from non-linear Luttinger liquid theory}
\label{app:pwrlaw}
In this section, we provide for reference the values of the power-law exponents of the spectral function for a homogeneous TG gas as obtained  using the mobile impurity or depleton model applied to the Lieb-Liniger Hamiltonian in  the limit of infinite interactions, as deduced from  Refs. \cite{Imambekov2009,Imambekov2012,Imambekov2008,Imambekov2009a,Campbell2017}.
The exponents are obtained in terms of phase shifts, which  can be written as\cite{Kamenev2009,Imambekov2009a}
\begin{equation}
\frac{\delta_\pm}{2 \pi} = \frac{1}{2}\left[\frac{1}{v(k)\mp v_s} \left( \frac{\sqrt{K}}{\pi} \frac{\partial \epsilon (k)}{\partial n} \pm \frac{1}{\sqrt{K}} \frac{k}{m}  \right) \mp \frac{1}{\sqrt{K}}\right], 
\end{equation}
where $K$ is the Luttinger parameter, $m$ is the mass of the particles, $\epsilon(k)$ is the dispersion of the excitation branch and $v(k)$ is the corresponding group velocity.
In the TG limit, obtained as the infinite interaction limit of the Lieb-Liniger model the value of the Luttinger parameter is  $K=1$. Correspondingly,  $v_s=k_F/m$, $v(k)=k/m$ and $n=k_F/\pi$. 
In order to calculate the exponents $\mu_A$ and $\mu_D$, respectively $\overline {\mu_+}$ and $\underline {\mu_-}$ of Ref.~\cite{Imambekov2009a}, we have to choose $\epsilon(k)=\epsilon_1(k)$, the Lieb-I curve of the main text. This results in $\frac{\delta_\pm}{2 \pi} =\frac{1}{2}$. Using Eqs.(9) and (10) of Ref.~\cite{Campbell2017} we then obtain
\begin{equation}
\mu_A = 1 - \frac{1}{2} \left(\frac{\delta_+-\delta_-}{2 \pi}  \right)^2 - \frac{1}{2} \left(\frac{\delta_+ +\delta_-}{2 \pi}  \right)^2 = \frac{1}{2}
\end{equation} 
\begin{equation}
\mu_D = 1 - \frac{1}{2} \left(2+\frac{\delta_+-\delta_-}{2 \pi}  \right)^2 - \frac{1}{2} \left(\frac{\delta_+ +\delta_-}{2 \pi}  \right)^2 = -\frac{3}{2}.
\end{equation} 
In order to calculate $\mu_B$ and $\mu_C$, respectively $\underline {\mu_+}$ and  $\overline {\mu_-}$ of Ref. \cite{Imambekov2009a}, we  choose $\epsilon(k)=-\epsilon_2(k)$, the Lieb-II curve of the main text, resulting in $\frac{\delta_\pm}{2 \pi} =-\frac{1}{2}$.  Again using  Eqs.(9) and (10) of Ref. \cite{Campbell2017} we obtain
\begin{equation}
\mu_C =  1 - \frac{1}{2} \left(\frac{\delta_+-\delta_-}{2 \pi}  \right)^2 - \frac{1}{2} \left(\frac{\delta_+ +\delta_-}{2 \pi}  \right)^2 = \frac{1}{2}
\end{equation} 
\begin{equation}
\mu_B =  1 - \frac{1}{2} \left(2+\frac{\delta_+-\delta_-}{2 \pi}  \right)^2 - \frac{1}{2} \left(\frac{\delta_+ +\delta_-}{2 \pi}  \right)^2 = -\frac{3}{2}.
\end{equation}
We finally remark that the presence of the lattice is expected to renormalize such exponents, except in the limit of very low filling.


\end{document}